\documentclass[aps,twocolumn,nofootinbib,groupedaddress,amsfonts,floatfix,prd]{revtex4-1}

\usepackage{graphicx,amsmath,amssymb,amstext}
\usepackage{amsbsy,amsfonts,amsthm}

\usepackage{bm}
\usepackage{subfigure}
\input epsf

\newcommand\lsim{\mathrel{\rlap{\lower4pt\hbox{\hskip1pt$\sim$}}
        \raise1pt\hbox{$<$}}}
\newcommand\gsim{\mathrel{\rlap{\lower4pt\hbox{\hskip1pt$\sim$}}
        \raise1pt\hbox{$>$}}}

\begin{document}
\title{How to Run Through Walls: Dynamics of Bubble and Soliton Collisions}

\author{John T. Giblin, Jr${}^{1,2}$}
\email[]{giblinj@kenyon.edu}
\author{Lam Hui${}^{3,4,5}$}
\email[]{lhui@astro.columbia.edu}
\author{Eugene A. Lim${}^3$}
\email[]{eugene.a.lim@gmail.com}
\author{I-Sheng Yang${}^3$}
\email[]{isheng.yang@gmail.com}
\affiliation{${}^1$Department of Physics, Kenyon College, Gambier, OH 43022, U. S. A.}
\affiliation{${}^2$Perimeter Institute for Theoretical Physics, 31 Caroline St N, Waterloo, ON N2L 2Y5, Canada}
\affiliation{${}^3$ISCAP and Department of Physics, Columbia University, New York 10027, U. S. A.}
\affiliation{${}^4$Institute for Advanced Study, Princeton, NJ 08540, U. S. A.}
\affiliation{${}^5$CCPP and Department of Physics, New York University, NY 10003, U. S. A.}

\date{\today}
\begin{abstract}

It has recently been shown in high resolution numerical simulations that 
relativistic collisions of bubbles in the context of a multi-vacua 
potential may lead to the creation of bubbles in a new vacuum.
%-- robust coherent field configurations which span a \emph{different} 
%part of the potential  are formed. 
%Such ``new'' walls will be long lived if the potential supports their existence. 
In this paper, we show that scalar fields with only potential interactions
behave like free fields during high-speed collisions; the kick
received by them in a collision can be deduced simply
by a linear superposition of the bubble wall profiles.
This process is equivalent to the scattering of solitons in $1+1$ dimensions.
We deduce an expression for the field excursion (shortly after a collision),
which is related simply to the field difference between the parent
and bubble vacua, i.e. contrary to expectations, the excursion cannot
be made arbitrarily large by raising the collision energy.
There is however a minimum energy threshold for this excursion to be
realized. We verify these predictions using a number of $3+1$ and $1+1$
numerical simulations. A rich phenomenology follows from these
collision induced excursions --- they provide a new mechanism for
scanning the landscape, they might end/begin inflation, and they might
constitute our very own big bang, leaving behind a potentially observable
anisotropy.
%In this paper, we show that this process is generic : during high-speed 
%collisions, the scalar fields behave like free fields and the dynamics 
%during the initial stages of the interaction can be approximated by 
%linear superposition, a process we call \emph{free passage}. 
%The superposition 
%Contrary to naive expectations, the maximal field excursion
%(shortly after a collision) cannot be made arbitrarily large by 
%increasing the collision energy, but is rather determined by this 
%superposition principle. We derive a minimum energy condition for
%this maximal excursion to be realized. A classical transition to
%a new vacuum follows if the excursion takes the fields to within
%the vacuum's basin of attraction.
%We verify these predictions using a number of $3+1$ bubbles and $1+1$ solitons 
%numerical simulations -- emphasizing that this process is 
%not confined to cosmological bubble walls.
%The collisionally produced vacuum region could accommodate our universe,
%implying interesting observational signals.

\end{abstract}

%\pacs{98.80.-k; 98.80.Es; 98.65.Dx; 98.62.Dm; 95.36.+x; 95.30.Sf; 04.50.Kd; 04.80.Cc}

%%04.80.Cc is for Experimental tests of gravitational theories
%%04.50.Kd is for Modified theories of gravity
%%95.30.Sf is for Relativity and gravitation
%%98.62.Gq is for Galactic halos
%%98.62.Dm is for kinematics, dynamics, and rotation of galaxies
%%98.65.Dx is for LSS
%%98.80.Bp is for Origin and formation of universe
%%98.80.Cq is for Particle-theory and field-theory models of early universe
%%95.35.+d is for Dark Matter
%%98.70.Vc is for Background radiations

\maketitle

\section{Introduction: \emph{What Happens Then?}}
\label{intro}

The recent interest in the phenomenology of the string landscape \cite{BP} has sparked a resurgence of interests in the physics of bubbles and 
bubble collisions.
% In particular, investigation in the latter often runs into a wall, figuratively 
% speaking, as the non-linear nature of such collisions force us to leave the 
% comfort of perturbative physics.
An understanding of such collisions promises much: a prediction of observational consequences of such collisions \cite{AguJoh09,LarLev09,FreKle09,AguJoh08,ChaKle08,ChaKle07,RubSak01}, an insight into the interaction of non-perturbative objects such as domain walls in general, a more complete treatment of the bubble counting measure \cite{EasLim05,GarSch05}, possibly a new method of scanning the landscape and many other wonderful things. As Coleman \cite{Col77} succinctly asked in his seminal paper on the fate of the false vacuum ``...bubble walls begin to collide. What happens then? Can such events be accommodated in the history of the early universe?'' 

The {\it kinematics} of wall collisions is well understood: provided one knows
the nature of the incoming and outgoing walls (if indeed walls are the only
by-products), the wall trajectories can be obtained by energy-momentum conservation
\cite{LanMae01,FreHor07, BouFre08a,AguJoh07}.
Our goal, on the other hand, is to understand the collision {\it dynamics}\cite{AxeKom99}:
how does the field configuration evolve through a collision?

\begin{figure*}[tb]
\subfigure[]
{\label{setup1}\includegraphics[width=.50\textwidth]{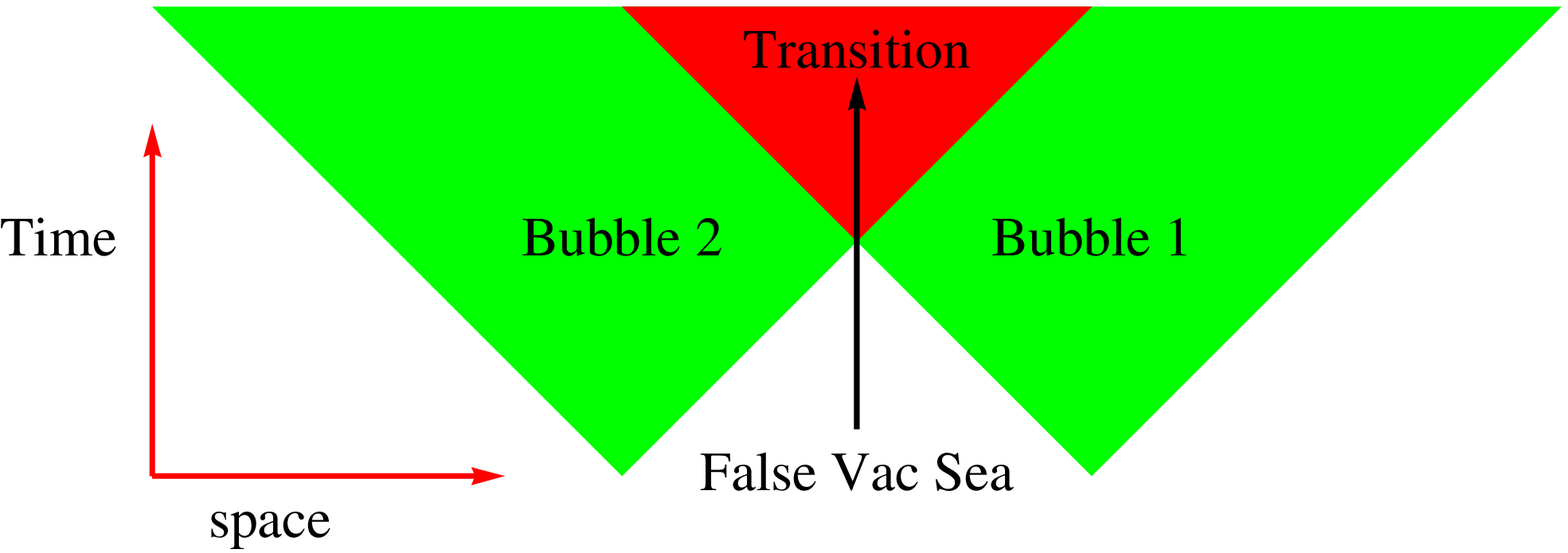}}
\hspace{0.1in}
\subfigure[]
{\label{setup2}\includegraphics[width=.40\textwidth]{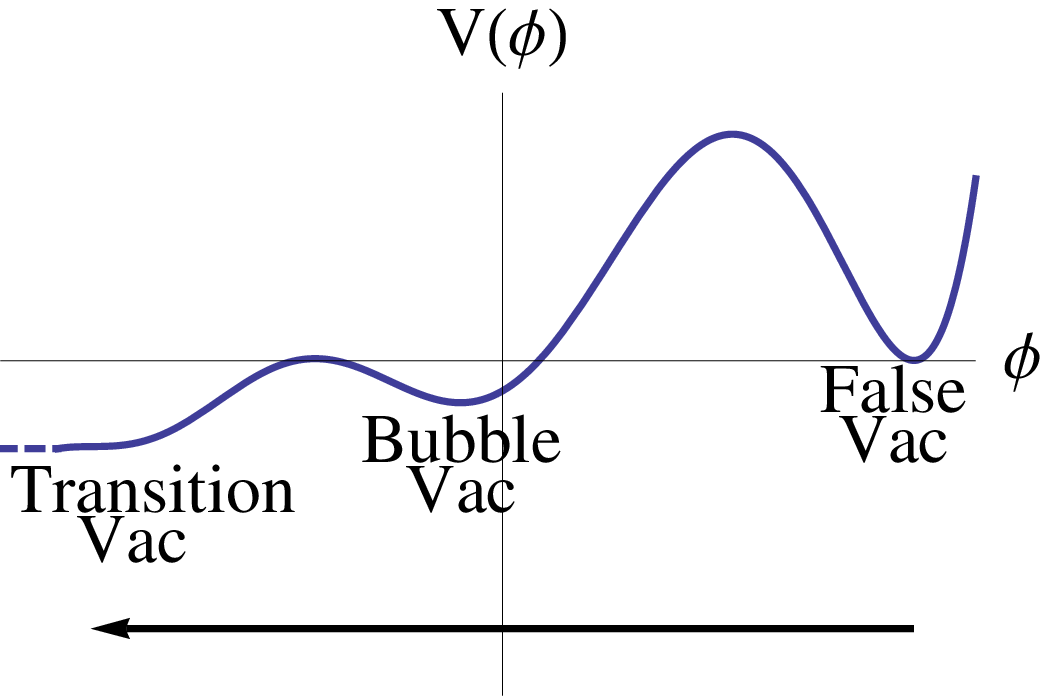}}
\caption{{\bf (a).} 
A space-time diagram showing the collision of
two bubbles embedded in a false vacuum sea, in a frame in which
the two bubbles nucleate at the same time.
We refer to the overlap between the bubbles as the transition region
-- here, the scalar field is generally kicked by the collision to
somewhere else in field space. 
{\bf (b).} A schematic diagram showing the potential for the
scalar field. 
The arrow denotes the scalar field kick as seen by an observer
following the black arrow in panel (a). This is the kick
immediately after a collision, which is realized as long as a certain
minimum condition is met. Where the field will roll to in the long
run depends on the precise layout of the landscape in the transition 
region. Note that only one bubble vacuum is shown. The other bubble
might inhabit the same or some other vacuum. Note also the
field space does not have to be one-dimensional.
}
\label{setup}
\end{figure*}

One fruitful approach to understanding the non-perturbative physics involved is numerical, namely lattice simulations of bubble wall collisions. This was
pioneered almost thirty years ago by Hawking, Moss and Stewart \cite{HawMos82a}. Despite their relatively (by today's computational standards) low resolution\footnote{See also \cite{KosTur92,KosTur91}, where lattice simulations were used to investigate the collisions of sub-horizon bubbles during
a first order electroweak phase transition.}, their 1+1 dimensional simulations discovered an interesting result that was not predicted by analytic methods: the collision of two true vacuum bubbles creates a relatively long live pocket of false vacuum which eventually collapse under differential pressure between the spacetime regions. This result was not well understood or exploited until recently.  In high-resolution, 3+1 dimensional, numerical simulations, it was shown that not only is this formation of false vacuum pocket a fairly robust effect, the right kind of potential can produce a new, stable bubble of a different vacuum (\cite{EasGib09}, henceforth EGHL). Furthermore, such \emph{classical transitions} can be
quite elastic --- coherent bubble walls form immediately after collisions with 
little dissipative losses in the form of scalar radiation.
EGHL also showed that whether such transition occurs is dependent on both the 
energy of the collision, and the height of the potential barrier between the progenitor bubble vacuum and the progeny bubble vacuum. 
The setup is illustrated in Fig. \ref{setup}.
%They argue that in a collision, gradient energy kept in the bubble wall is 
%efficiently converted into kinetic energy in field space, providing a large 
%enough kick to force the field to go over the potential barrier into the progeny vacuum.

In this paper, we wish to extend the work of EGHL, beginning with
the question: what determines the size of the
collision induced excursion? 
By this we mean the field excursion at $x=0$ 
(the location of the collision),
from immediately before to immediately after the collision
(Fig. \ref{setup}). 
Naively, one might think that the excursion depends on the collision
energy --- the higher the energy, the further the 
scalar field could go --- as if the kinetic energy stored in
the bubble walls can be tapped to initiate motions in field space.
This kind of reasoning proves to be misleading, however.

Let us try to gain some insight by simplifying the problem.
Recall that bubble walls accelerate as the bubbles grow in size. 
A high speed collision therefore also means a collision between two large
bubbles. In this limit, we can treat the collision event as one
between two planar walls, in other words it is a 1+1 dimensional problem.
The collision of two bubble walls is thus equivalent to the collision
of two kink solitons.
Much is known about the scattering of solitons in for instance
the famous sine-Gordon model (Appendix \ref{sinegordon}). 
We will be able to deduce some general model independent results
in the relativistic collision limit, including the case of multiple scalar
fields.
And we will find that the collision induced field excursion is
in fact bounded, and given by a simple expression involving the
field difference between the false and the bubble vacua
(\S \ref{freepassage}). There is a minimum energy threshold
that must be reached for this maximal field excursion to 
be realized, which is derived in \S \ref{transition}. 
These analytic predictions are verified with numerical experiments
in \S \ref{simulations}. 
We will digress a bit in \S \ref{multiple} to consider
cases where the excursion takes the field through multiple
local minima. 

These collision induced field excursions open up a rich set of
questions, including: How do they impact inflation? 
What are the observational signatures if we reside in the
transition region (Fig. \ref{setup})? What happens when there
are multiple collisions? 
We will explore these questions in \S \ref{sect:discussion}.
For the most part, we focus in this paper on scalar field
models with canonical kinetic terms and potential interactions,
and work in flat space. 
Relaxing these assumptions will be discussed in \S \ref{sect:discussion}
as well.

\section{The Free Passage Approximation -- Implication for
the Collision-induced Field Excursion}
\label{freepassage}

We are interested in solutions to
the scalar field equation,
\begin{eqnarray}
\label{boxphi}
\Box \phi = {\partial V \over \partial \phi} \, ,
\end{eqnarray}
which contain bubble/domain wall configurations.
Here, $V(\phi)$ is a potential with two or more metastable minima. 
In particular, we are interested in how the domain/bubble walls interact
in a collision. 
It should be emphasized that while we are primarily motivated
by cosmological applications, much of our discussion carries over
to the collision of domain or solitonic walls in broader contexts.

For the moment, we focus on a single scalar field $\phi$.
The generalization to multiple fields will be discussed below. 
We work in Minkowski space and defer a discussion of gravity
to \S \ref{sect:discussion}.
Some of our numerical simulations do have an expanding background,
but the backreaction of the scalar field on the geometry
is not properly taken into account. 

Let us follow the strategy laid out in \S \ref{intro}, and
take the high speed/large bubble limit, in which case
the collision problem becomes effectively $1+1$ dimensional:
%Let us further simplify the problem by focusing on the
%collision of large bubbles. Cosmological bubbles 
%generally accelerate as
%they grow in size, and so large bubbles also mean high speed.
%When the bubble radii are large compared their thickness,
%the collision can be approximated as one between two
%planar walls, i.e. the problem becomes 1+1 dimensional:
\begin{eqnarray} \label{eqn:1+1dEOM}
-\partial_t^2 \phi + \partial_x^2 \phi = {\partial V \over \partial \phi} \, ,
\end{eqnarray}
where $t$ is time and $x$ labels the axis of collision.

\begin{figure*}[tb]
\subfigure[]
{\label{schematicpotential1}\includegraphics[width=.45\textwidth]{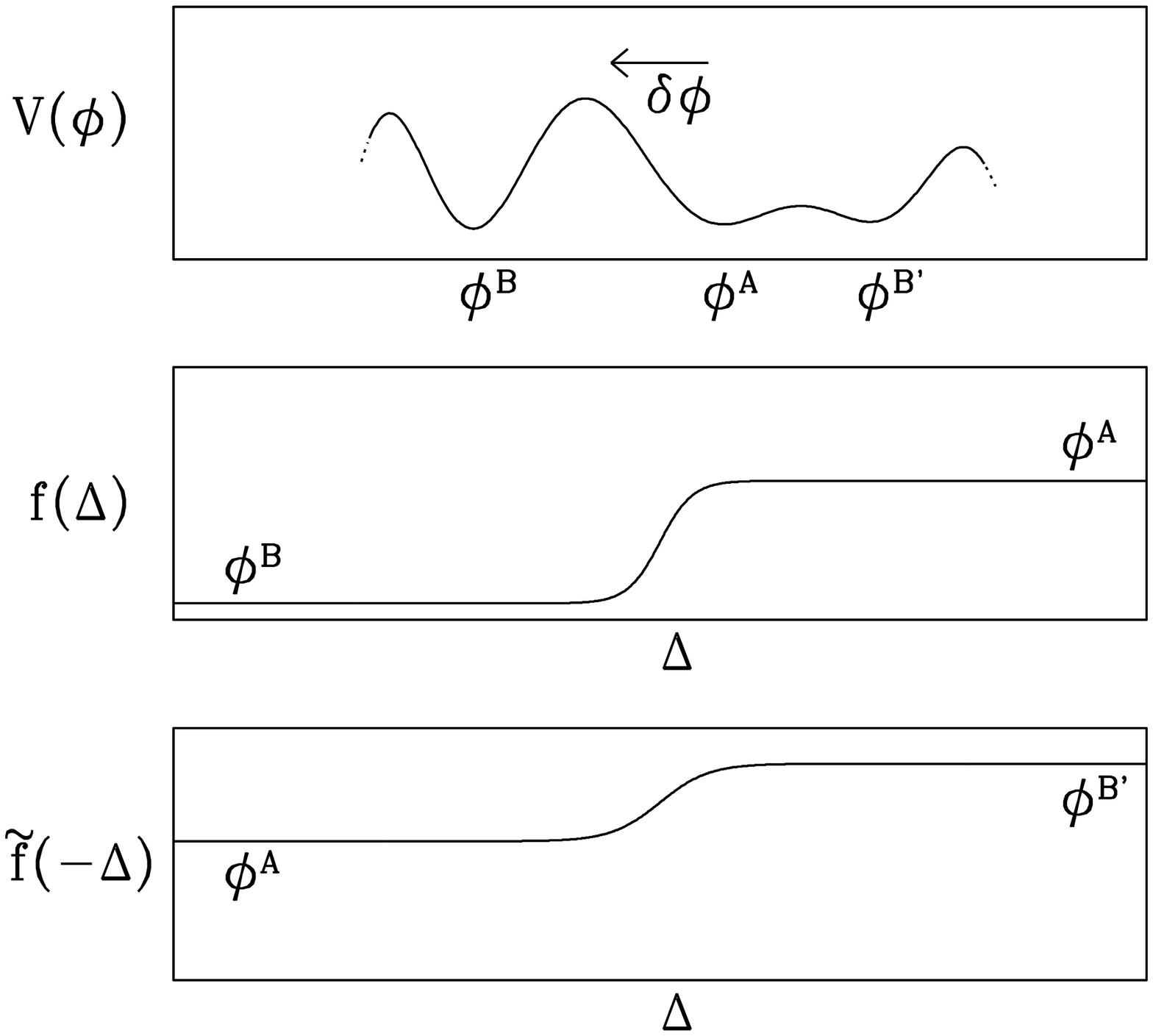}}
\hspace{0.1in}
\subfigure[]
{\label{schematicphi1}\includegraphics[width=.45\textwidth]{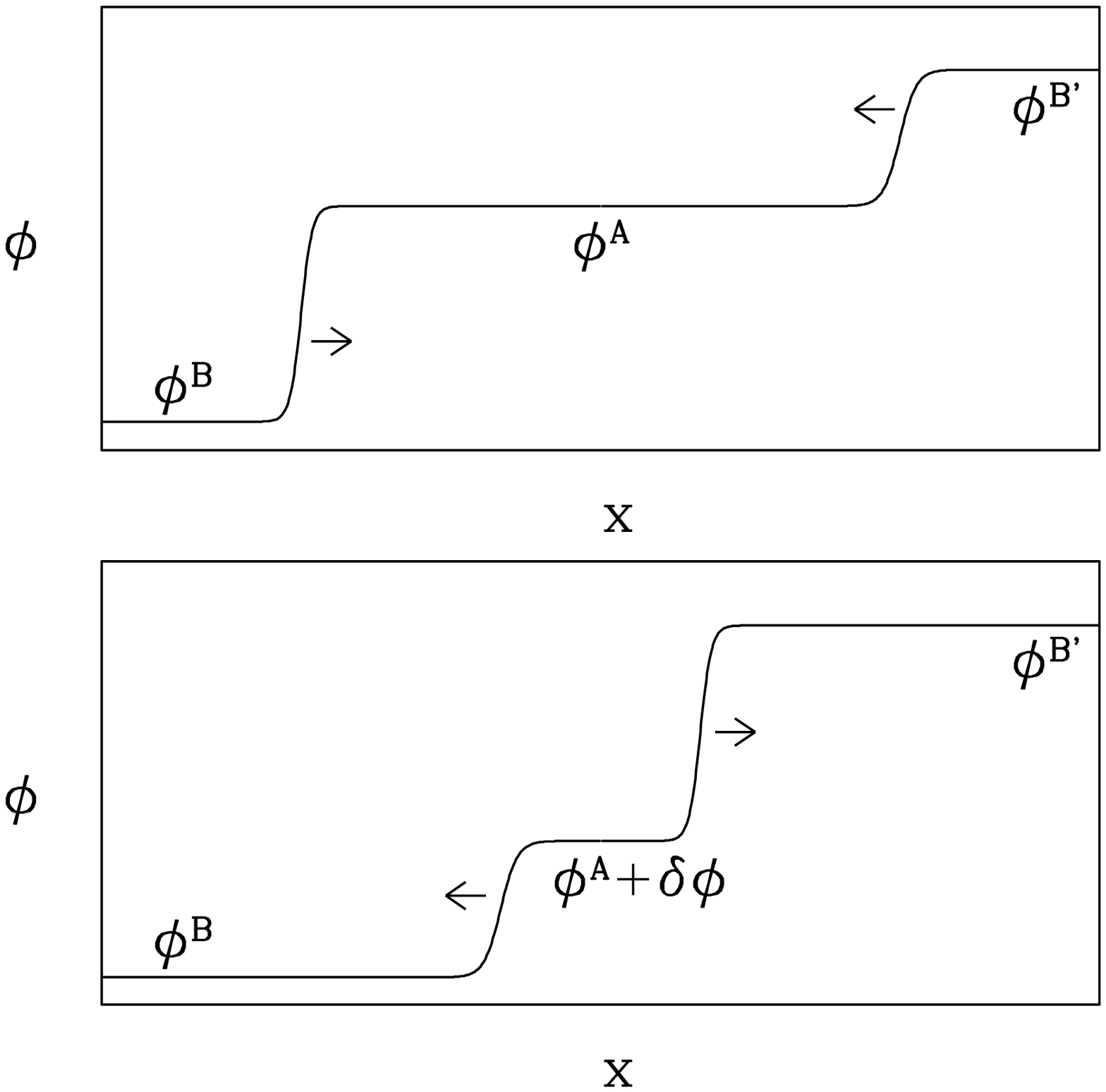}}
\caption{{\bf (a).} 
The top panel displays part of a landscape of local minima of the potential 
$V$ of
some scalar field $\phi$. 
$\delta\phi$ is the kick induced by collision.
The middle panel shows the solitonic profile 
that interpolates
between $\phi^B$ and $\phi^A$. The bottom panel shows the solitonic profile
that interpolates between $\phi^{B'}$ and $\phi^A$. 
{\bf (b).} The top panel depicts the scalar field $\phi$ as a function of position $x$ at some time $t < 0$ in the far past, before the two solitons collide.
These two solitons can be thought of as part of bubble walls: the bubble
on the left has an interior scalar field value of $\phi^B$; the bubble on
the right has an interior scalar value of $\phi^{B'}$; $\phi^A$ in
between is the false/parent vacuum.
The bottom panel shows the scalar profile soon after the solitons collide. 
What is assumed is the free passage of waveforms (Eq. (\ref{phisum})):
the top panel comes from summing $f$ to the {\it left} of $\tilde f$, the
bottom from summing $f$ to the {\it right} of $\tilde f$. 
Note how the field value
in the collision region shifts from $\phi^A$ pre-collision
to $\phi^A + \delta\phi$ post-collision, with $\delta\phi = \phi^B + \phi^{B'}
- 2\phi^A$. In other words, the outgoing objects are {\it different}
from the incoming solitons: they interpolate between different field values.
}
\label{schematic1}
\end{figure*}

We are interested in a potential of the sort schematically
shown in Fig. \ref{schematicpotential1}. 
The `false' (or parent) vacuum is denoted by $\phi^A$, and
the the two bubble vacua are $\phi^B$ and $\phi^{B'}$ respectively.
In other words, the collision of interest is between two walls,
one spanning from $\phi^A$ to $\phi^B$, and the other from $\phi^A$ to 
$\phi^{B'}$. We will treat these vacua as roughly degenerate,
so that the wall speed is nearly constant. This conflicts somewhat
with our picture of generally accelerating bubble walls, but the
degeneracy assumption is not really crucial --- much of our discussion
carries over to the non-degenerate case with the speed $u$ below
taken to be the wall velocity just prior to collision.

In the rest frame of the respective solitonic walls, the wall profiles
are denoted by $f$ and $\tilde f$ (Fig. \ref{schematicpotential1}),
i.e. they satisfy the equations:
\begin{eqnarray}
\label{fdef}
{d^2 f(\Delta) \over d \Delta^2} = {\partial V(f) \over \partial f}
\quad , \quad
{d^2 \tilde f(\Delta) \over d \Delta^2} = {\partial V(\tilde f) \over \partial \tilde f} \, ,
\end{eqnarray}
with the respective boundary conditions
$f(\Delta \rightarrow \infty) = \phi^A$,
$f(\Delta \rightarrow -\infty) = \phi^B$, and
$\tilde f(\Delta \rightarrow \infty)
= \phi^A$, $\tilde f(\Delta \rightarrow -\infty) = \phi^{B'}$.

Boosting to a frame in which the two solitons collide
at equal and opposite speed $u$, the wall profiles 
as a function of space and time are:
\begin{eqnarray}
f \left({x-ut \over \sqrt{1-u^2}} \right)
\quad , \quad 
\tilde f \left(- {x + ut \over \sqrt{1 - u^2}} \right)
\, .
\end{eqnarray}

%\footnote{One can always boost to a frame where the two bubbles nucleate at the same time, assuming the nucleation events are space-like separated.}

%Since we have assumed that the local minima at $\phi^A$, $\phi^B$
%and $\phi^{B'}$ are degenerate, the velocity $u$ is a constant
%\footnote{When they are not degenerate, $u$ accelerates with time. Our main
%results below should still hold approximately as long as one uses
%$u$ as the velocity just prior to collision.}.
Here, $f$ represents the right going soliton and
$\tilde f$ represents the left going one. 
%The setup is illustrated
%in Fig. \ref{schematicpotential1}: $\phi^A$ is the field value of the 
%parent vacuum sea, and $\phi^B$ and $\phi^{B'}$ are the respective
%field values interior to the two bubbles.
When the solitons are far apart, say at $t\rightarrow -\infty$, 
the solitons only interact very weakly with each other, and hence  we can 
represent the \emph{total} scalar field solution $\phi$ as
a linear superposition of the two solutions $f$ and $\tilde{f}$
\begin{eqnarray}
\label{phisum}
\phi(t, x) = f \left( {x-ut \over \sqrt{1-u^2}} \right) + 
\tilde f \left( - {x + ut \over \sqrt{1 - u^2}} \right) - \phi^A \, .
\end{eqnarray}
The constant shift of $-\phi^A$ is necessary so that at $t < 0$
(prior to collision), the scalar field $\phi$ takes the 
(false/parent vacuum) value $\phi^A$ between the two walls, and
the field to the far left and far right takes the values
$\phi^B$ and $\phi^{B'}$ respectively
(upper panel of Fig. \ref{schematicphi1}).

While it is easy to see that Eq. (\ref{phisum}) is a good approximation
when the two solitons are far apart, the key insight in understanding soliton collisions is that \emph{Eq. (\ref{phisum}) remains a good approximation
even during and slightly past the collision time $t=0$, for sufficiently energetic collisions i.e. $u\rightarrow 1$.}  
In this limit, both the spatial and time gradient terms
dominate over the potential term in Eq. (\ref{eqn:1+1dEOM})
at the walls (the wall profiles effectively become step functions):
\begin{eqnarray}
\label{lineareqt}
| \partial_x^2 \phi|~,~| \partial_t^2 \phi| \gg \left| \frac{\partial V}{\partial \phi}\right|~,~-\partial_t^2 \phi + \partial_x^2 \phi \sim 0 \, .
\end{eqnarray}
Meanwhile, away from the walls, the scalar field
sits roughly at the corresponding local
minimum where the derivative of the potential is also small.

A superposition of the two waveforms $f$ and $\tilde f$
as in Eq. (\ref{phisum}) should therefore be a good solution to the linear
wave equation (\ref{lineareqt}), as long as $u \sim 1$. This should
hold even post-collision.
(Let us postpone for the moment a discussion of when 
this approximation Eq. [\ref{phisum}] ceases to be accurate.)
Taking this seriously gives a very interesting
prediction: around $x = 0$ where the collision happens, 
$\phi = \phi^A$ before collision, whereas $\phi = \phi^B + \phi^{B'} - \phi^A$
after collision.
In other words, compare the upper and lower panels of
Fig. \ref{schematicphi1}: the pre-collision configuration is
obtained by superimposing $f$ to the left of $\tilde f$, whereas
the post-collision configuration comes from superimposing $f$ to the
right of $\tilde f$ (taking care to include the constant
shift $-\phi^A$ as in Eq. [\ref{phisum}]).
In other words, the two wall profiles 
(or waveforms) \emph{freely pass through} each other, creating a region where the scalar field takes a value
that is neither the original parent vacuum value $\phi^A$, nor the bubble interior values
$\phi^B$ or $\phi^{B'}$. We say that the collision produces a \emph{kick}
to the scalar field of
\begin{eqnarray}
\label{1Dkick}
\delta \phi = \phi^B + \phi^{B'} - 2 \phi^A \, .
\end{eqnarray}
This is illustrated  in Fig. \ref{schematicphi1}. We refer to
this observation as the \emph{free passage approximation}. 

It is worth noting that even as the two profiles or 
waveforms $f$ and $\tilde f$
freely pass through each other, the nature of the
solitons has actually changed: while the incoming solitons interpolate between
$\phi^A$ and $\phi^B$ (and between
$\phi^A$ and $\phi^{B'}$), the outgoing objects
interpolate between $\phi^A + \delta\phi$ and $\phi^B$ (and between
$\phi^A + \delta\phi$ and $\phi^{B'}$). 
Indeed, the outgoing objects strictly speaking are not even solitons, since
$\phi^A + \delta\phi$ is not in general a stationary point of the potential (see Fig. \ref{schematicpotential1}).
One could think of the collision process as a scattering event.
The free passage of waveforms is reminiscent of the impulse approximation
used in high energy scattering.

\begin{figure*}[tb]
\subfigure[]
{\label{schematicpotential2}\includegraphics[width=.45\textwidth]{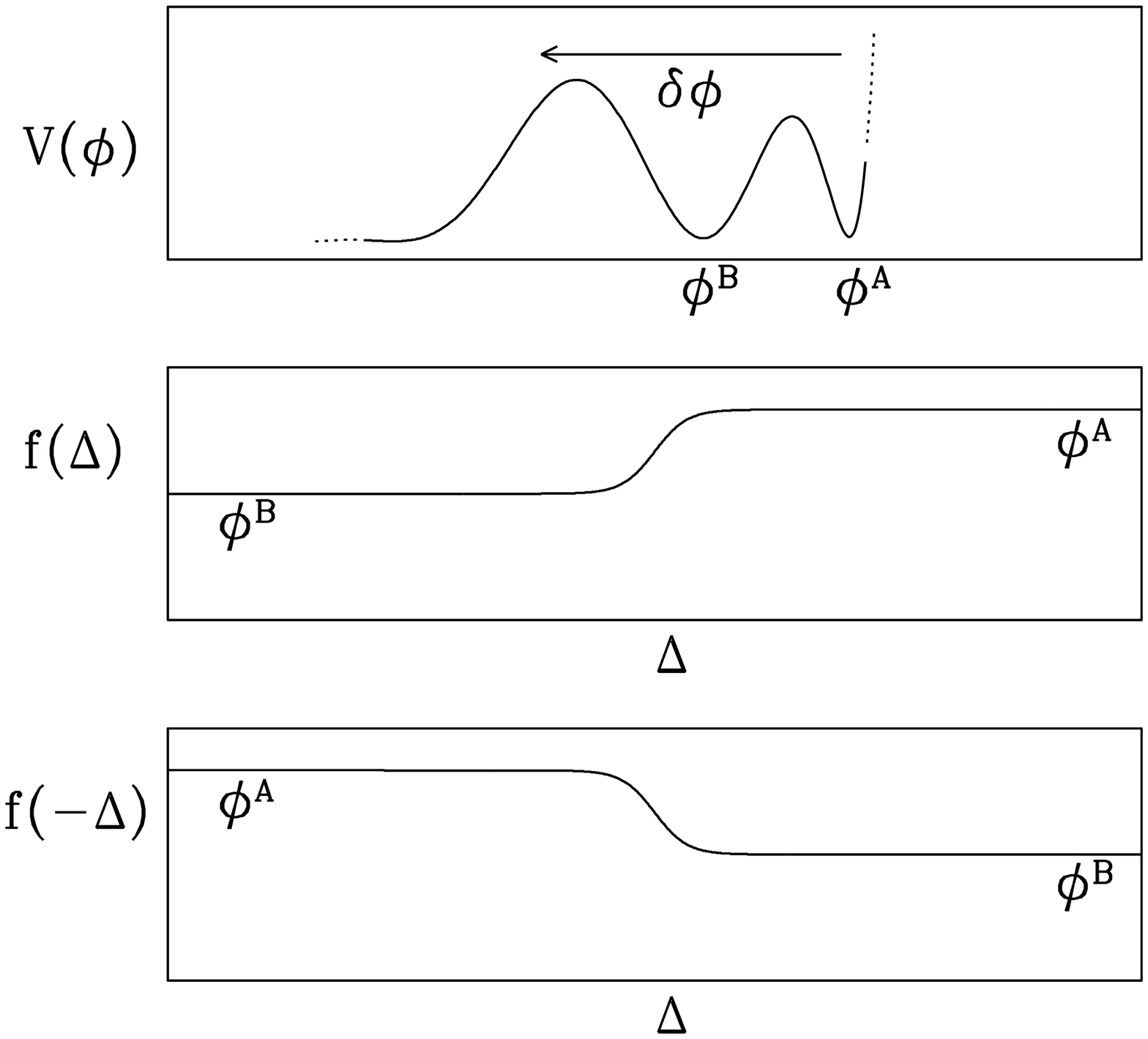}}
\hspace{0.1in}
\subfigure[]
{\label{schematicphi2}\includegraphics[width=.45\textwidth]{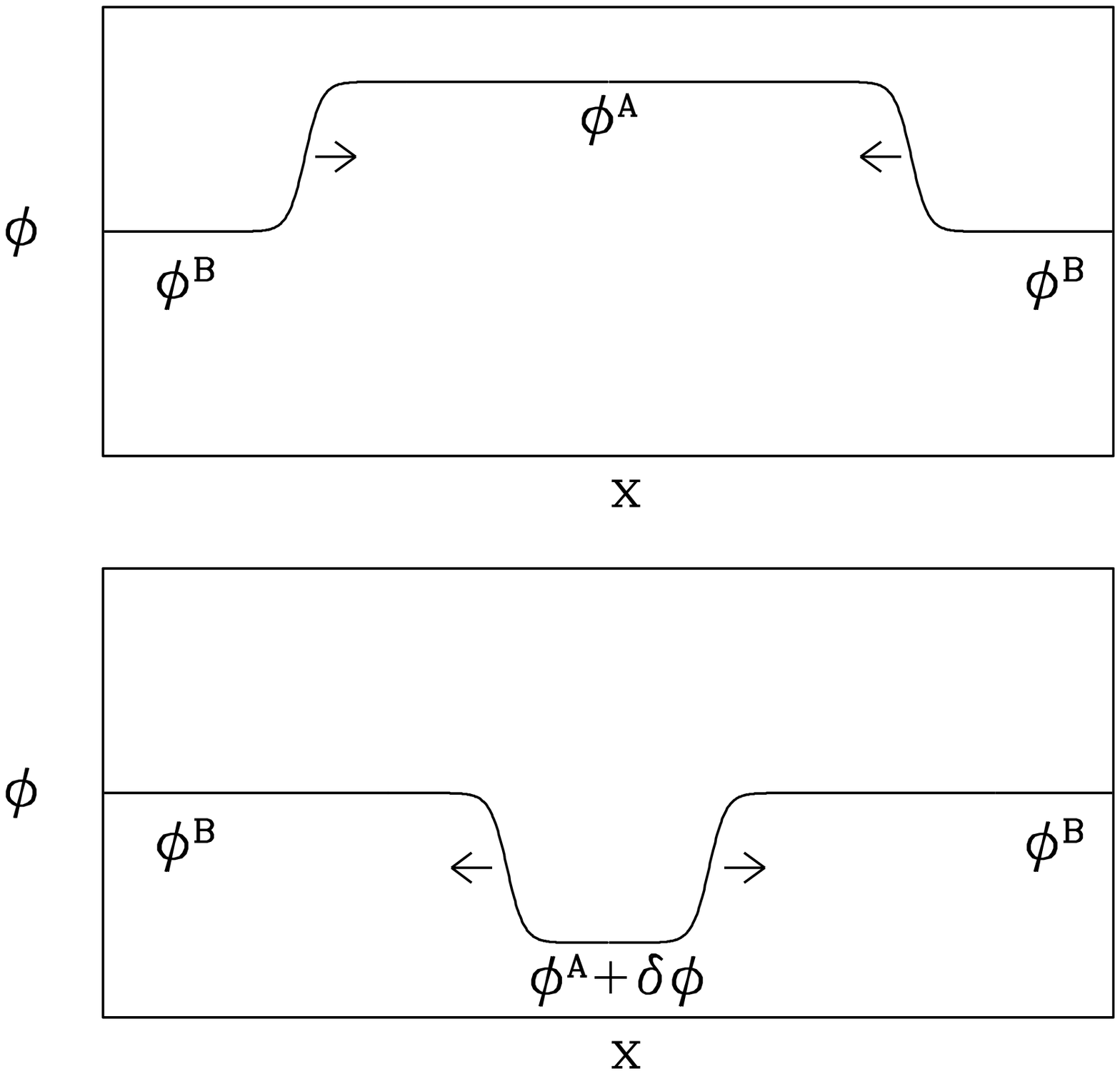}}
\caption{Analog of Fig. \ref{schematic1} for
$\phi^{B'} = \phi^{B}$,
such that the kick $\delta \phi = 2 (\phi^B - \phi^A)$.
As before, the incoming configuration in the top panel of
{\bf (b)} corresponds to summing $f(\Delta)$ to the left of 
$f(-\Delta)$, while the outgoing configuration in the bottom
panel of {\bf (b)} corresponds to summing $f(\Delta)$ to the
right of $f(-\Delta)$ (more precisely: the free passage
solution in Eq. (\ref{phisum})).
Physically, one can think of this
as the annihilation of a soliton and an anti-soliton
giving rise to a {\it different} pair of objects.
From the point of view of bubble collisions, 
we have here two $\phi^B$-bubbles (immersed in
the parent vacuum sea $\phi^A$) colliding with each other,
triggering a new bubble forming at $\phi^A + \delta\phi$ (at least
momentarily).
}
\label{schematic2}
\end{figure*}

The free passage approximation should break down soon
after free passage itself. 
%This is because the field value
%of $\phi^A + \delta\phi$ in the collision region is not a stationary
%point of the potential in general
%(see Fig. \ref{schematicpotential1}). 
As illustrated in the lower
panel of Fig. \ref{schematicphi1}, there is a region between
the outgoing objects where the field gradient is small while
the derivative of the potential is non-negligible
(i.e. $\phi^A + \delta\phi$ is not a stationary point of the potential).
Eq. (\ref{lineareqt})
therefore ceases to be a good approximation, and one must account for
the effect of the potential.
%(which in fact is acting even during the early moments of the 
%collision even if it is subdominant).
%In short, immediately after (and in fact during)
%the kick $\delta\phi$, the potential becomes important.

Under what condition then, is the free passage kick
$\delta\phi = \phi^B + \phi^{B'} - 2 \phi^A$ actually realized?
In the subsequent sections,
we will estimate, and verify with numerical calculations, 
a minimum collision energy/velocity in order for the 
free passage kick to be successfully executed.
If the collision is not sufficiently energetic,
the scalar field can still make a fleeting excursion of a similar size,
but it quickly retreats back to the false or bubble vacua thereafter.
We will refer to this as an unsuccessful or failed kick. Anticipating
a bit numerical experiments we will run, 
the distinction between a successful and a failed kick is illustrated
in Fig. \ref{marchmovie} \& \ref{retreatmovie}.
%%In other words, under what condition would the field in
%%the collision region be able to reach this predicted excursion
%%and indeed march past it? In numerical calculations that will
%%be shown below, we find that if the collisional energy (or $u$) is
%%not sufficiently high, the field would barely make this predicted
%%excursion and quickly retreat. 
%We will derive in the next section a minimum energy condition that
%delineates march versus retreat -- here we will discuss the qualitative 
%predictions of the free passage approximation.
%\footnote{Anticipating a bit numerical experiments we will perform
%later, the distinction between march and retreat is illustrated
%in Fig. \ref{marchmovie} \& \ref{retreatmovie}.}

It is important to emphasize that our primary goal here is
to work out the field excursion shortly after a collision, rather
than its long term behavior.
As noted above, the free passage approximation generally breaks
down soon after a collision (even if the kick is successful),
and therefore cannot be used to predict the long term outcome of 
the field dynamics.
The short term outcome, however, does have long term implications:
if the free passage kick fails, the field $\phi$
would inevitably roll back to the false or bubble vacua; 
if the free passage kick succeeds, the field's long
term evolution would be driven by whatever the basin of attraction
it happens to be in after the kick. If $\phi^A + \delta\phi$ is
in the basin of attraction of some new vacuum, then we can have
the formation of a new vacuum bubble.

%By march/retreat, we 
%are referring to
%the short term, as opposed to long term, behavior of the field. 
%If in the short term $\phi$ keeps marching forward (in direction of
%the kick), 
%its long term behavior (where it ultimately
%lands) will be determined by
%whatever basin of attraction it happens to be in.
%If in the short term $\phi$ retreats,
%(as we will see below,
%this means in practice the
%field barely reaches the theoretical maximum 
%of $\delta\phi = \phi^B + \phi^{B'} - 2 \phi^A$),
%then $\phi$ rolls back to the false/parent or bubble vacua.

This is why the case illustrated in Fig. \ref{schematic1}
is in a sense not so interesting, for no matter whether
in the short term the free passage kick succeeds or not,
in the long run it would not roll to any new vacuum.
%roll to $\phi^A$, $\phi^B$ or $\phi^{B'}$ (depending
%on the exact shape of the potential) i.e.
%there is no creation of a bubble inhabiting a new vacuum.
A more interesting example
is one where $\phi^B = \phi^{B'}$ i.e. colliding two bubbles inhabiting
the same vacuum. This is illustrated in Fig. \ref{schematic2}.

What happens in this case is that the collision induced kick
$\delta \phi = 2(\phi^B - \phi^A)$ could lead
to the formation of a new bubble that is at neither
the parent vacuum $\phi^A$ nor the bubble vacuum $\phi^B$.
Indeed, in the example depicted in
%the top panel of Fig. \ref{schematicpotential2},
%it appears $\delta\phi$ is sufficient for the field to
%overcome a barrier to the left of $\phi^B$ -- we will
%call this the ``new'' barrier to distinguish it from the
%original barrier between $\phi^A$ and $\phi^B$.
%Suppose for the moment, contrary to the situation depicted in Fig. 2, 
%$\delta\phi = 2(\phi^B - \phi^A)$ is actually not sufficiently large to 
%climb past
%the new barrier to the left of $\phi^B$. In that case, the
%field in the collision region would eventually retreat back to
%$\phi^B$ or even $\phi^A$ -- this is because both the potential
%and the gradient energy favor a retreat.
Fig. \ref{schematic2}, $\delta\phi$ is sufficiently
large that the kick takes the field over a new barrier (at the far left).
Provided the collision
satisfies a minimum energy condition to be 
derived in the next section, we would have an interesting
outcome:  the creation of a bubble of a new vacuum in the collision region. 

%Since the free passage solution already roughly possess the ``shape'' of the new solitonic wall, modulo any future roll-down into the new minimum, relaxation of this temporary free passage solution into an actual soliton solution will dissipate some energy in form of the emission of  scalar radiation. The amount of dissipation naturally depends on the details of the potential and may be large, however, \emph{the free passage approximation predicts that the generic outcome of a successful transition is the formation of a coherent soliton and not the annihilation of the two incoming walls.} This is the crucial reason why classical bubbles are nucleated in the collision of cosmological bubbles that was first observed numerically in EGHL.

It is worth noting that the free passage approximation predicts
the walls retain their integrity through a collision (even as
they change character: a wall that interpolates between
$\phi^B$ and $\phi^A$ become one that interpolates between
$\phi^B$ and $\phi^A + \delta\phi$). 
It is perhaps surprising that the collision does not result primarily
in dissipation into scalar radiation instead.
The reason is because at a high collision speed, the (potential) interactions
are negligible and the field is essentially free (Eq. [\ref{lineareqt}]). 
As we will see in \S \ref{simulations}, this simple reasoning
appears to be borne out by numerical simulations. 
There will be some radiative loss, but the amount tends to be small.
The potential interactions are of course important after the free passage kick.
For instance, in the example depicted in Fig. \ref{schematic2}, 
after a successful kick, the scalar field will roll towards whatever new vacuum that
exists on the far left. The outgoing walls
will adjust their profiles, as the region between them relaxes towards
this new vacuum. 

Conversely, if the collision does not satisfy the minimum energy
condition, then the field excursion barely reaches
$\delta\phi = 2(\phi^B - \phi^A)$, and eventually retreats back to
the false/parent or bubble vacua, $\phi^A$ or $\phi^B$. As we will see,
this can happen even if, {\it momentarily}, the field has crossed
the new potential barrier to the left. 
The retreat is due to tendency of the spatial
gradient force to counteract the (free-passage implied) field excursion. 
%Whether the field marches onward or retreats
%depends on a competition between potential
%energy and gradient energy -- the former pushes $\phi$ onward
%(if the derivative of the potential has the appropriate sign),
%while the latter favors a retreat. 
In the language of solitons, a retreat (i.e. an unsuccessful kick) 
means the outgoing pair of (soliton-like) 
objects cannot overcome their mutual attraction, a phenomenon
we will observe in numerical simulations (Fig. \ref{retreatmovie}).

%The above discussion begs an interesting question : can we determine from the potential and in particular the field-space distance between the vacua, make a prediction whether such classical transition can occur given an ultra relativistic collision? The answer is yes. 

%Suppose for the moment, contrary to the situation depicted in Fig. 2, 
%$\delta\phi = 2(\phi^B - \phi^A)$ is actually not sufficiently large to 
%climb past
%the new barrier to the left of $\phi^B$. In that case, the
%field in the collision region would eventually retreat back to
%$\phi^B$ or even $\phi^A$ -- this is because both the potential
%and the gradient energy favor a retreat.
Lastly, to cover the range of possible outcomes, consider a situation where
contrary to Fig. \ref{schematic2}, the kick $\delta\phi = 2(\phi^B - \phi^A)$
is not large enough to take the scalar field over the new barrier to the left.
In this case, even if the kick were successful, the scalar field
would have to roll back to $\phi^B$ or even $\phi^A$ eventually. 
More quantitatively, one can see that at the collision point ($x = 0$),
\begin{equation}
\frac{\partial^2\phi}{\partial t^2} = \frac{\partial^2\phi}{\partial x^2} -\frac{\partial V}{\partial\phi} > 0
\end{equation}
after free passage (recalling that the free passage approximation
predicts vanishing space and time derivatives at $x=0$ after a collision). 
Therefore, the field retreats back towards $\phi^B$ or $\phi^A$.

In summary, the free passage kick of $\delta\phi = \phi^B + \phi^{B'} - 2\phi^A$,
or $\delta\phi = 2(\phi^B - \phi^A)$ if the bubbles inhabit the same vacuum, 
constitutes a maximal field excursion shortly after a collision.

\begin{figure}[tb]
\centerline{\epsfxsize=9cm\epsffile{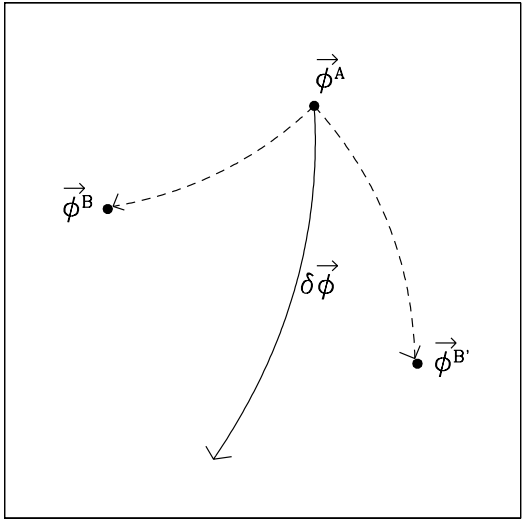}}
\caption{A schematic diagram {\it in field space},
illustrating the collision induced
kick in a landscape of a multi-dimensional scalar field
$\vec\phi$.
$\vec\phi^A$ denotes the false/parent vacuum value. Embedded inside
the false vacuum sea are two bubbles, one inhabiting
the local minimum at $\vec\phi^B$, the other at $\vec\phi^{B'}$.
When these two bubbles collide, the field in the collision region
would undergo an excursion $\delta\vec\phi$ (at least momentarily) away from
$\vec\phi^A$, where $\delta\vec\phi$ is simply the vector
sum of $\vec\phi^B - \vec\phi^A$ and $\vec\phi^{B'}-\vec\phi^A$.
The dashed lines denote the bubble wall profile from vacuum $A$ to
vacuum $B$ and vacuum $B'$ respectively. 
The solid line is a linear superposition of these two wall profiles,
and illustrates the field excursion upon bubble collision
predicted by free-passage (Eq. [\ref{vectorFP}]).
}
\label{multiscalarkick}
\end{figure}

{\bf Generalization to multiple fields.}
Let us conclude this section by pointing out
that all of our arguments above translate in a straightforward manner
to the case of multiple scalar fields, i.e. the equation of motion is
\begin{eqnarray}
\label{boxphimultiple}
\Box \vec\phi = {\partial V \over \partial \vec\phi} \, .
\end{eqnarray}                        
Suppose we have a landscape of a multiple-dimensional scalar $\vec \phi$.
Suppose further that we have a parent or false vacuum sea at $\vec \phi^A$,
within which there are two bubbles, one at $\vec \phi^B$
and one at $\vec \phi^{B'}$. When these two bubbles collide,
the free passage approximation tells us that
the scalar field should receive a kick in the collision region
from $\vec \phi^A$ to $\vec \phi^A + \delta\vec\phi$, where
\begin{eqnarray}
\label{multidimkick}
\delta\vec\phi = \vec\phi^B + \vec\phi^{B'} - 2\vec\phi^A \, .
\end{eqnarray}
The situation is illustrated in Fig. \ref{multiscalarkick}.
This prediction for $\delta\vec\phi$ should be accurate
momentarily after a high speed collision. 
Note that the linear superposition (Eq. [\ref{phisum}]) actually
says more than this: not only is there a vector sum rule
governing the end point of the excursion Eq. (\ref{multidimkick}),
the colliding wall profiles, which are not 
necessarily straight lines in field space,
also add to give the excursion profile.

Let us close by noting two crucial assumptions: 
that there are only potential interactions,
and that the multi-scalar kinetic term (in the action)
is diagonalized and canonical. 
We will discuss deviations from these in \S \ref{sect:discussion}.
%The details of multifield models will be presented in an upcoming publication.

%by whatever basin of
%attraction $\vec \phi^A + \delta\vec\phi$ happens to be in.

\section{A Minimum Energy Condition for a Successful Kick}
\label{transition}

Let us consider under what condition the free passage approximation
is accurate, i.e. that the scalar field successfully realizes
the field excursion $\delta \phi = 2(\phi^B - \phi^A)$
in the example depicted in Fig. \ref{schematic2}.
More general situations, such as that depicted
in Fig. \ref{multiscalarkick}, will be considered later. 

The field configuration can be expressed as
\begin{eqnarray}
\phi(t, x) = \phi_{\rm fp} (t, x) + \sigma (t, x) \, ,
\end{eqnarray}
where $\phi_{\rm fp}$ is the free-passage solution
which is a linear superposition of right-going soliton
and left-going anti-soliton Eq. (\ref{phisum}):
\begin{eqnarray}
\phi_{\rm fp} (t, x) = f \left( {x-ut \over \sqrt{1-u^2}} \right) + 
{f} \left( - {x + ut \over \sqrt{1 - u^2}} \right) - \phi^A \, 
\end{eqnarray}
and $\sigma$ represents the deviation from the actual solution from the free passage solution. 
The free passage approximation is accurate only when 
$\sigma$ is small compared to, say, the free passage kick itself i.e.
$\sigma/\delta \phi \ll 1$. 
We will use the shorthand $f_R \equiv f([x-ut]/\sqrt{1-u^2})$
and $f_L \equiv f(-[x+ut]/\sqrt{1-u^2})$.

Using Eq. (\ref{fdef}), we can rewrite the $\phi$ equation of motion as an equation for $\sigma$
\begin{eqnarray}
\label{eomsigma}
\Box \sigma
= {\partial V \over \partial \phi} \Big |_{\phi = \phi_{\rm fp}
+ \sigma} 
- {\partial V \over \partial \phi} \Big |_{\phi = f_R} - {\partial V \over \partial \phi} \Big |_{\phi = f_L} \, .
\end{eqnarray}
The question is: under what condition does $\sigma$
remain small after the right-going soliton and left-going anti-soliton have passed
through each other? 

A formal solution to Eq. (\ref{eomsigma}) is
%Notes 09 VII p. 112
\begin{eqnarray}
\label{sigmasolution}
\sigma(t, x) = - \int_{-\infty}^t dt' \int^{x + t - t'}_{x - (t - t')}
dx' g(t', x')
\end{eqnarray}
with the kernel function  $g(t', x')$ 
\begin{eqnarray}
\label{gdef}
g \equiv {\partial V \over \partial \phi} \Big |_{\phi = \phi_{\rm fp}
+ \sigma} 
- {\partial V \over \partial \phi} \Big |_{\phi = f_R} - {\partial V \over \partial \phi} \Big |_{\phi = f_L} \, .
\end{eqnarray}
We estimate this integral by ignoring $\sigma$ in the integrand
(first term in $g$) -- the free-passage approximation is self-consistent
if the resulting estimate for $\sigma$ is indeed small.
We are most interested in its value at $x=0$, and by making a change of variables $X' \equiv x'/\sqrt{1-u^2}$ and $T' \equiv ut'/\sqrt{1-u^2}$ we can rewrite it as follows to show its explicit dependence on the velocity $u$:
%Notes 09 VII p. 128
\begin{eqnarray}
\label{sigmaapprox}
&& \sigma(t,0) = - {1 - u^2 \over u} \int_{-\infty}^{ut\over
\sqrt{1-u^2}} dT' \int_{-(t-t')\over\sqrt{1-u^2}}^{t-t'\over\sqrt{1-u^2}} dX'
\nonumber \\
&& \quad \left[ {\partial V \over \partial \phi} \Big |_{\phi = f_R + f_L - \phi^A}
- {\partial V \over \partial \phi} \Big |_{\phi = f_R} - {\partial V \over \partial \phi} \Big |_{\phi = f_L} \right] \, .
\end{eqnarray}
%where $f = f(X' - T')$ and $\tilde{f} = \tilde{f}(-X' - T')$. 
Note that as $u\rightarrow 1$, $\sigma\rightarrow 0$ confirming our assertion that free passage is the right approximation in the relativistic limit.
The time $t$ we are interested  in is when the
solitons have just passed through each other, i.e.
$ut /\sqrt{1 - u^2} = \mu^{-1}$, where $\mu^{-1}$ is the rest-frame
thickness of the bubble walls. The thickness can be estimated by
$\mu^2 \sim |\partial^2 V/\partial \phi^2|$ evaluated at the barrier
that separates the parent and the bubble vacua.
The integrand in Eq. (\ref{sigmaapprox}) vanishes when
the two incoming solitons are far apart, and is dominated by
when $T' \sim \pm \mu^{-1}$. A reasonable estimate can be obtained
by focusing on what happens post-collision, when the integrand
is dominated by the first term $\partial V/\partial \phi$
at $\phi = 2\phi^B - \phi^A$. Therefore, we have
\begin{eqnarray}
\label{sigmaapprox2}
\sigma \sim - {1 - u^2 \over u^2} {1\over \mu^2} 
{\partial V \over \partial \phi} \Big |_{\phi = 2\phi^B - \phi^A}
\, .
\end{eqnarray}
As we argued above, self-consistency of the free passage approximation demands
that $|\sigma| \ll |\delta\phi| = |2(\phi^B - \phi^A)|$. 
To be concrete, we demand that for the field to continue marching onward after the free-passage kick, the collision velocity must satisfy
\begin{eqnarray}
\label{gammacondition}
\gamma^2 = {1 \over 1-u^2} \gsim 1 + {\alpha^{-1}\over |\phi^B - \phi^A|} {1\over \mu^2}
\Big | {\partial V \over \partial \phi} \Big |_{\phi = 2\phi^B - \phi^A}
\, 
\end{eqnarray} 
where the efficiency factor $\alpha^{-1}\approx 5.5$ is determined numerically. 
This efficiency factor tells us how much smaller
$|\sigma|$ has to be compared to $|\delta\phi|$ for free passage
to be a good approximation. 
%Alternatively, one could think of
%$1-\alpha$ as the fraction of the maximum kick $\delta \phi$ that is achieved 
%before free passage breaks down and the field begins to feel the potential. 

A collision that satisfies the condition
Eq. (\ref{gammacondition}) can successfully realize the
free passage kick. This does not by itself mean there
is a transition to a new vacuum --- for that to happen,
the kick must take the field to within the basin of attraction
of a new vacuum.
Conversely,  
%Note that this does \emph{not} signal that a transition has occur -- it merely 
%means that the field continues marching onwards after the initial free passage kick, 
%with its future evolution determined by the exact shape of the potential.
%{\bf Eugene: please check that the fudge factor here is chosen
%correctly to match your simulations.}
%The choice is made because not only does Eq. (\ref{gammacondition})
%provide a useful self-consistency condition for free passage, it
%also describes, according to numerical calculations,
%fairly accurately the condition for the 
%field excursion to be successfully executed
%i.e. the field continues marching onward after the free-passage
%kick. 
%In other words, collisions 
%satisfying Eq. (\ref{gammacondition}) give rise to a field excursion
%$\delta\phi = 2(\phi^B - \phi^A)$ that matches the free passage prediction,
%{\it and} the field subsequently marches beyond that excursion point.
%Conversely, 
collisions that do not satisfy Eq. (\ref{gammacondition})
produces a fleeting field excursion that barely makes $\delta\phi = 2(\phi^B - \phi^A)$,
and the field beats a retreat soon after.
%{\bf TO DO: Perhaps this is a good place to show some figures or simulations?}

There exists an important caveat to Eq. (\ref{gammacondition}): 
if $\partial V/\partial \phi$ at $\phi = 2\phi^B - \phi^A$ 
vanishes, this minimum energy condition must be re-evaluated.
What we have done in writing down Eq. (\ref{gammacondition})
(and Eq. [\ref{sigmaapprox2}]) was to approximate the full integrand
in Eq. (\ref{sigmaapprox}) by one single term. This is obviously a simplification
that needs to be fixed if the term we have chosen happens to vanish.
This happens, for instance, when $\phi = 2\phi^B - \phi^A$ is another local
minimum of the potential.
%then the second term of the equation is zero and hence the condition breaks down -- 
%it is easy to see that by raising the peak of the barrier while keeping the  
%locations of the two minima fixed  will increase the minimum $\gamma$ 
%required\footnote{We explicitly verified this numerically.}.
%In this case, terms we have ignored in deducing Eq. (\ref{gammacondition})
%should be taken into consideration
\footnote{
%A particularly simple case: $\phi = 2\phi^B - \phi^A$ 
%happens to be a local minimum of the potential and is degenerate with 
%the parent and bubble vacua. 
In this case, if the different minima are degenerate, 
the outgoing pair of objects are then true 
solitons, and Eq. (\ref{gammacondition}) can be replaced by:
%Notes 09 VI p. 120
$1/(1-u^2) \gsim m_{\rm out}^2/m_{\rm in}^2$, 
where $m_{\rm  out}$ and $m_{\rm in}$ are the rest-mass (or tension)
of the outgoing and incoming solitons, i.e.
$m = \int d\phi \sqrt{2V}$ with the limits of integration ranging over
the appropriate vacua ($V$ being set to zero at $\phi^B$ and $\phi^A$).
Energy conservation implies 
$m_{\rm out}^2 / m_{\rm in}^2 = (1 - v^2)/(1 - u^2)$, where
$v$ is the outgoing velocity and $u$ is the incoming one.
Demanding $v \le 1$ yields the inequality.
Allowing for radiation losses strengthens it
\cite{EasGib09}.
}

Finally, an intriguing counter example to the above caveat is the well known sine-Gordon 
soliton, where
the free-passage field excursion is successfully realized no matter what
value $u > 0$ takes \cite{ScoChu73}. In this case, the condition Eq. (\ref{gammacondition}) 
is both trivial and exact.  A brief summary is provided in Appendix \ref{sinegordon}.

{\bf Generalization to multiple fields.}
The above reasoning translates straightforwardly to the case
of multiple scalar fields, with the equation of motion
(\ref{boxphimultiple}).
The solution can be expressed as
$\vec \phi = \vec \phi_{\rm fp} + \vec \sigma$,
with the free-passage solution:
\begin{eqnarray}
\label{vectorFP}
\vec \phi_{\rm fp} (t, x) = 
\vec f \left( {x-ut \over \sqrt{1-u^2}} \right) + 
\vec {\tilde f} \left( - {x + ut \over \sqrt{1 - u^2}} \right) - 
\vec \phi^A \, .
\end{eqnarray}
where $\vec f$ interpolates between $\vec \phi^B$ and $\vec \phi^A$,
and $\vec {\tilde f}$ interpolates between $\vec \phi^{B'}$
and $\vec \phi^A$. 
Straightforward analogs of Eqs. (\ref{sigmasolution}),
(\ref{sigmaapprox}) and (\ref{sigmaapprox2}) can be written down.
The net free-passage field excursion is given by Eq. (\ref{multidimkick}).
A crude self-consistency condition for a successful kick is therefore:
\begin{eqnarray}
\label{gammacondition2}
{1 \over 1-u^2} \gsim 1 + {\alpha^{-1} \over |(\vec \phi^B + \vec \phi^{B'})/2 - 
\vec \phi^A |} {1\over \mu^2}
\Big | {\partial V \over \partial \vec \phi} \Big | \, ,
\end{eqnarray}
where the derivative is evaluated at
${\vec 
\phi = \vec \phi^B + \vec \phi^{B'}
- \vec \phi^A}$, and
$\mu^{-1}$ is the rest-frame thickness of the thicker wall.
The same caveat about a vanishing potential gradient at
the excursion point for the single field applies here as well.

To close this section, let us
 state the {\it transition condition} for the creation of a bubble
inhabiting a new (i.e. neither parent nor original bubble) vacuum via 
a collision: {\it the collision energy must be
high enough to satisfy Eq. (\ref{gammacondition2}), and the free-passage
excursion must take the field to within the basin of attraction of
a new vacuum.
}

\section{Numerical Solutions}
%{A toy model in 3+1}
\label{simulations}

\begin{figure}[tb]\centerline{\epsfxsize=9cm\epsffile{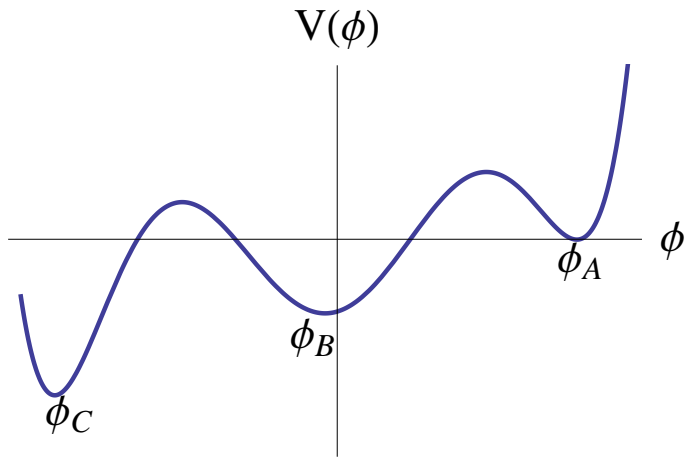}}
\caption{The potential Eq. (\ref{fulltoypot})  with three minima at $\phi_A$,$\phi_B$ and $\phi_C$. 
For the $3 + 1$ simulations, we nucleate two bubbles of $\phi_B$ from a sea of $\phi_A$ at the distance $d = 2\gamma R_0$ where $R_0$ is the initial bubble radius and $\gamma$ is the Lorentz factor at collision.}
\label{fig:Tompot}
\end{figure}

\begin{figure*}[tb]
\subfigure[]
{\label{march1}\includegraphics[width=.25\textwidth]{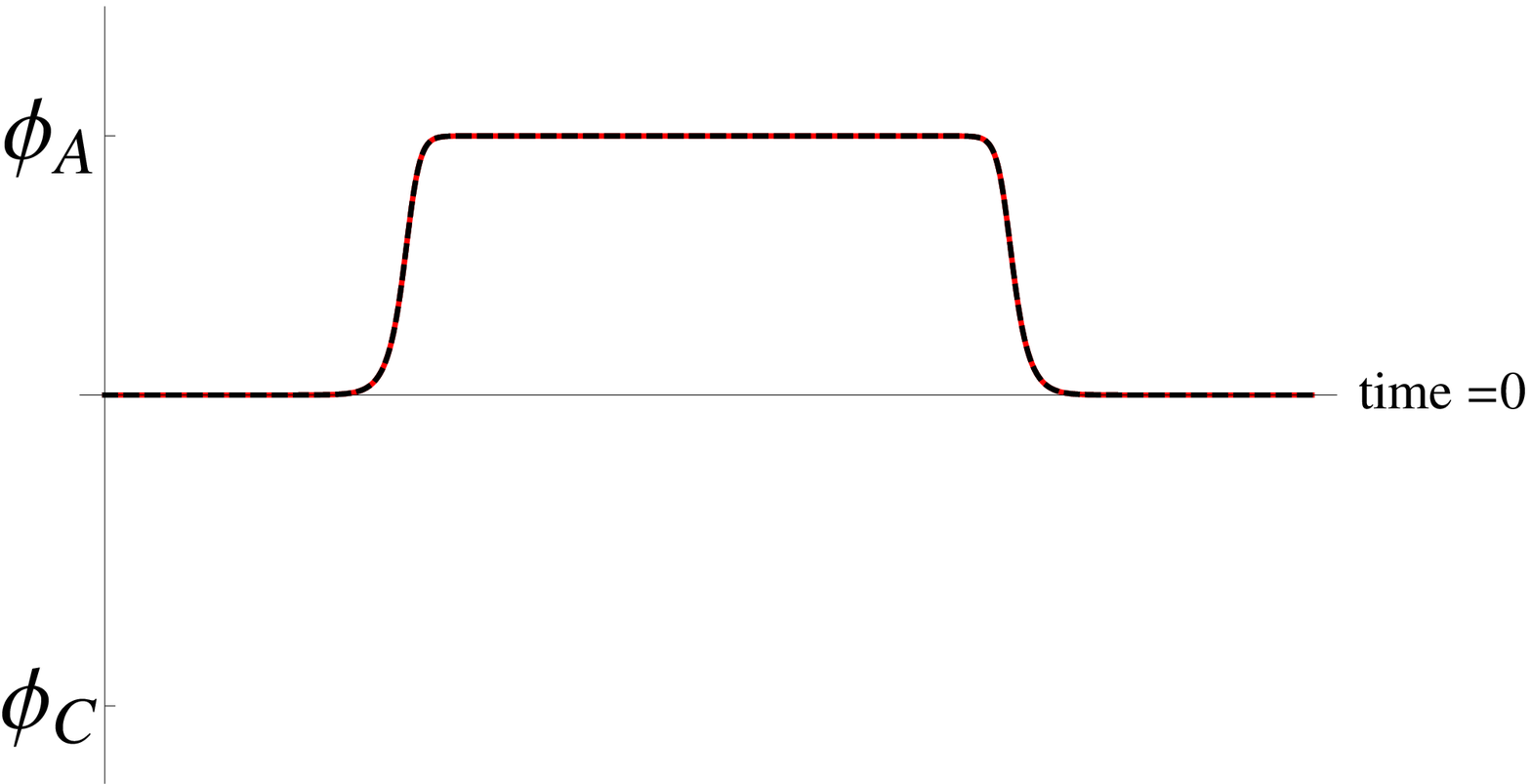}}
%\vspace{0.01in}
%
\subfigure[]
{\label{march2}\includegraphics[width=.25\textwidth]{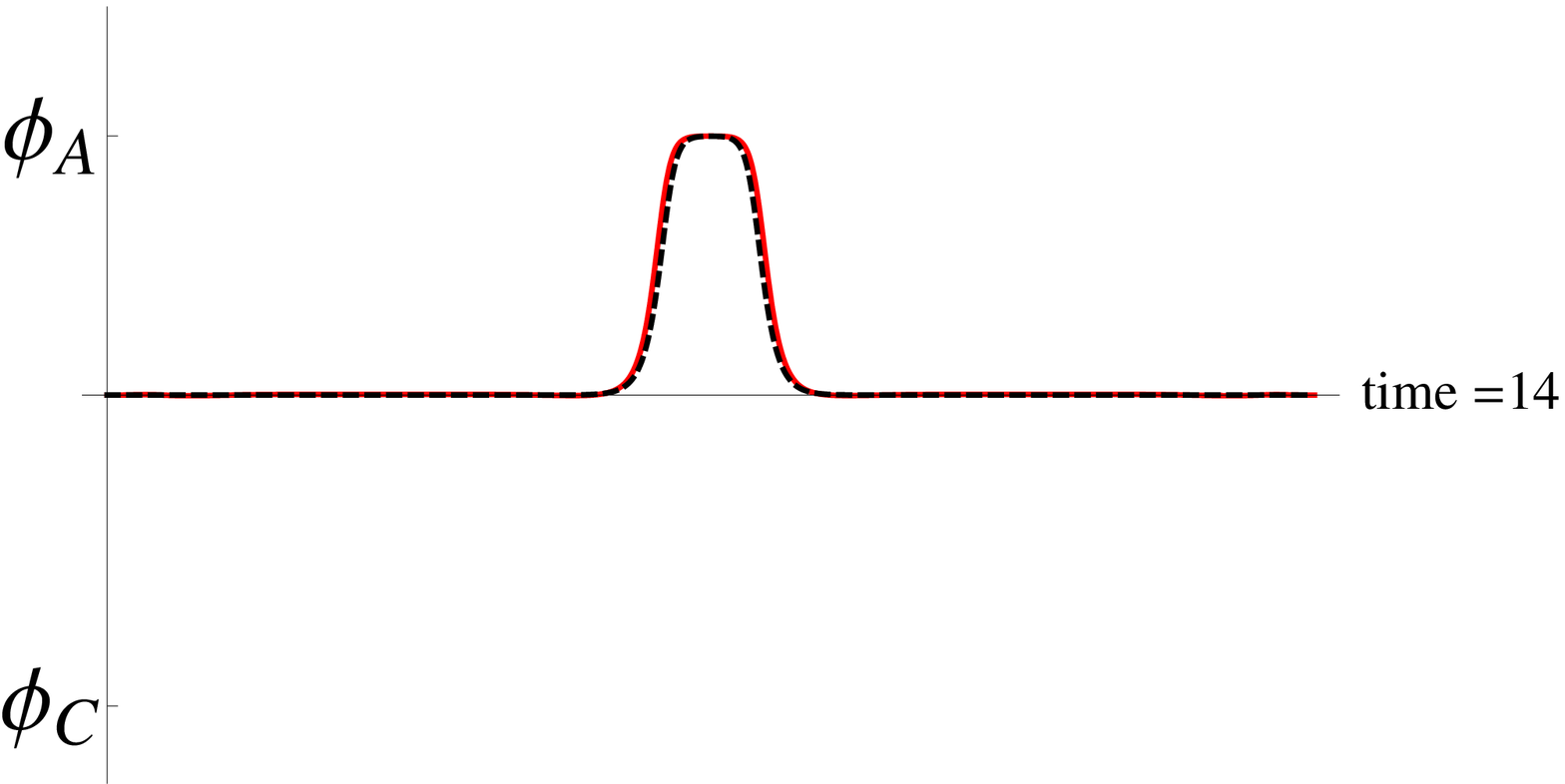}}
%\hspace{0.01in}
%
\subfigure[]
{\label{march3}\includegraphics[width=.25\textwidth]{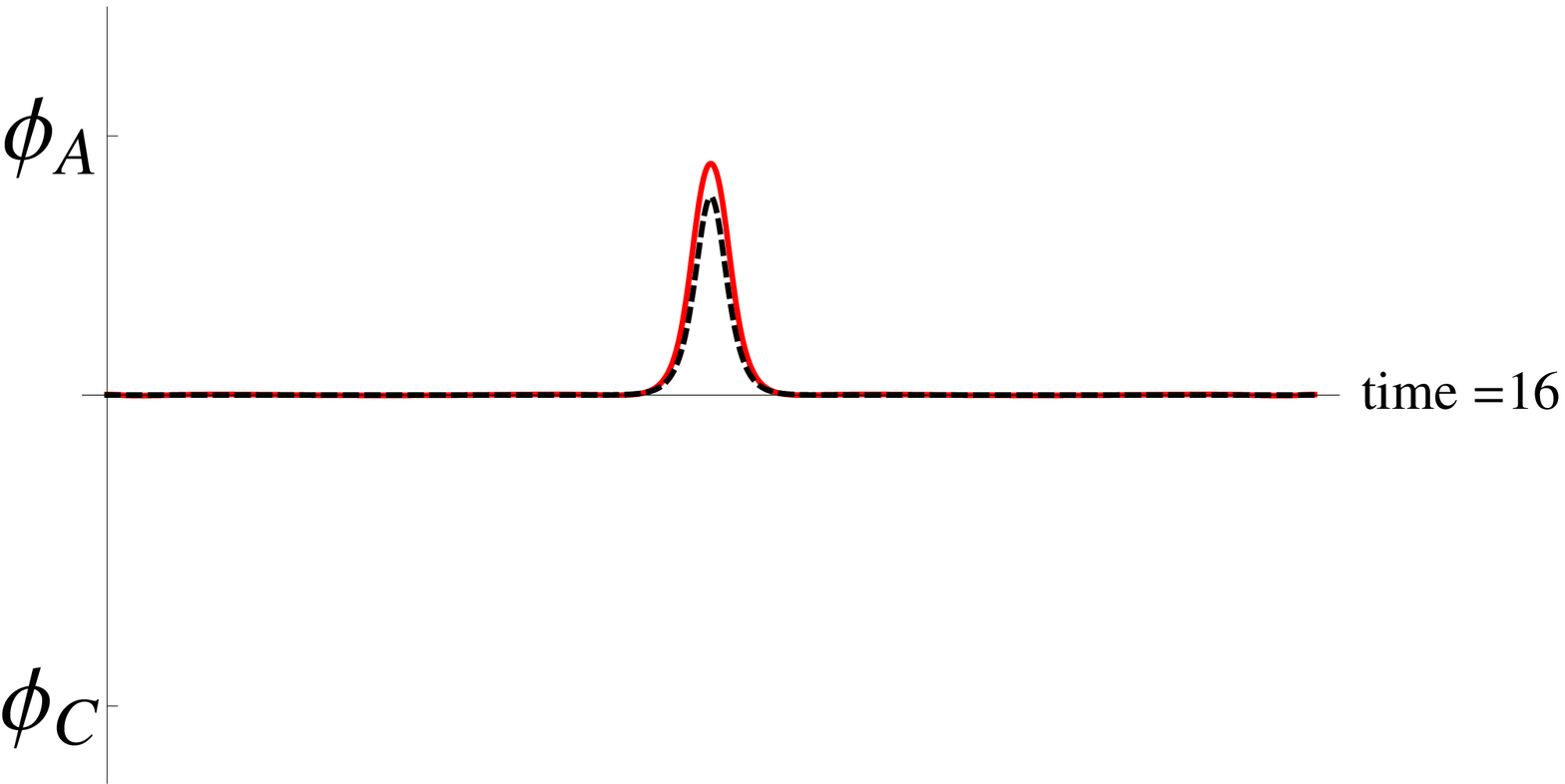}}
\subfigure[]
{\label{march4}\includegraphics[width=.25\textwidth]{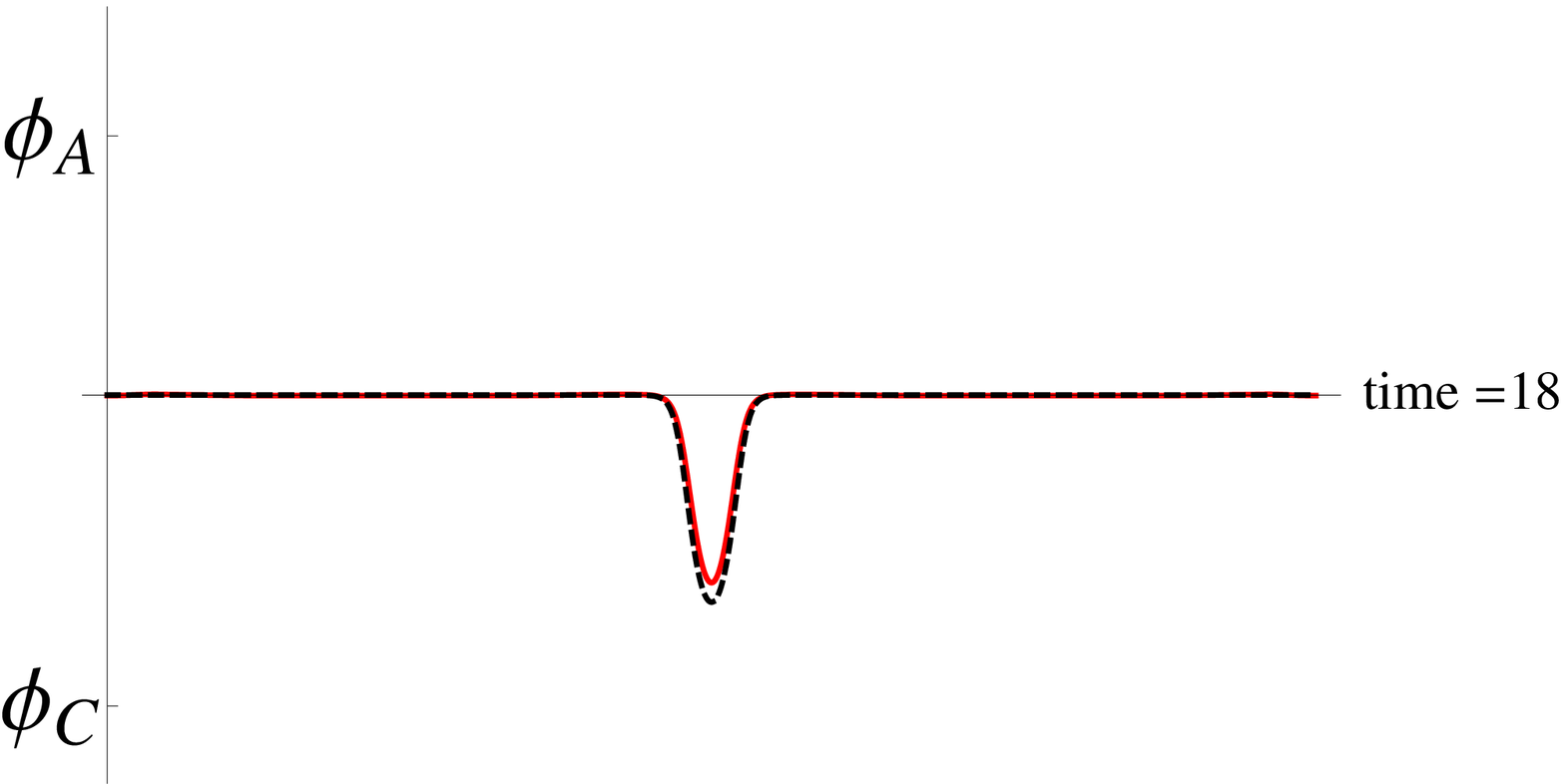}}
\subfigure[]
{\label{march5}\includegraphics[width=.25\textwidth]{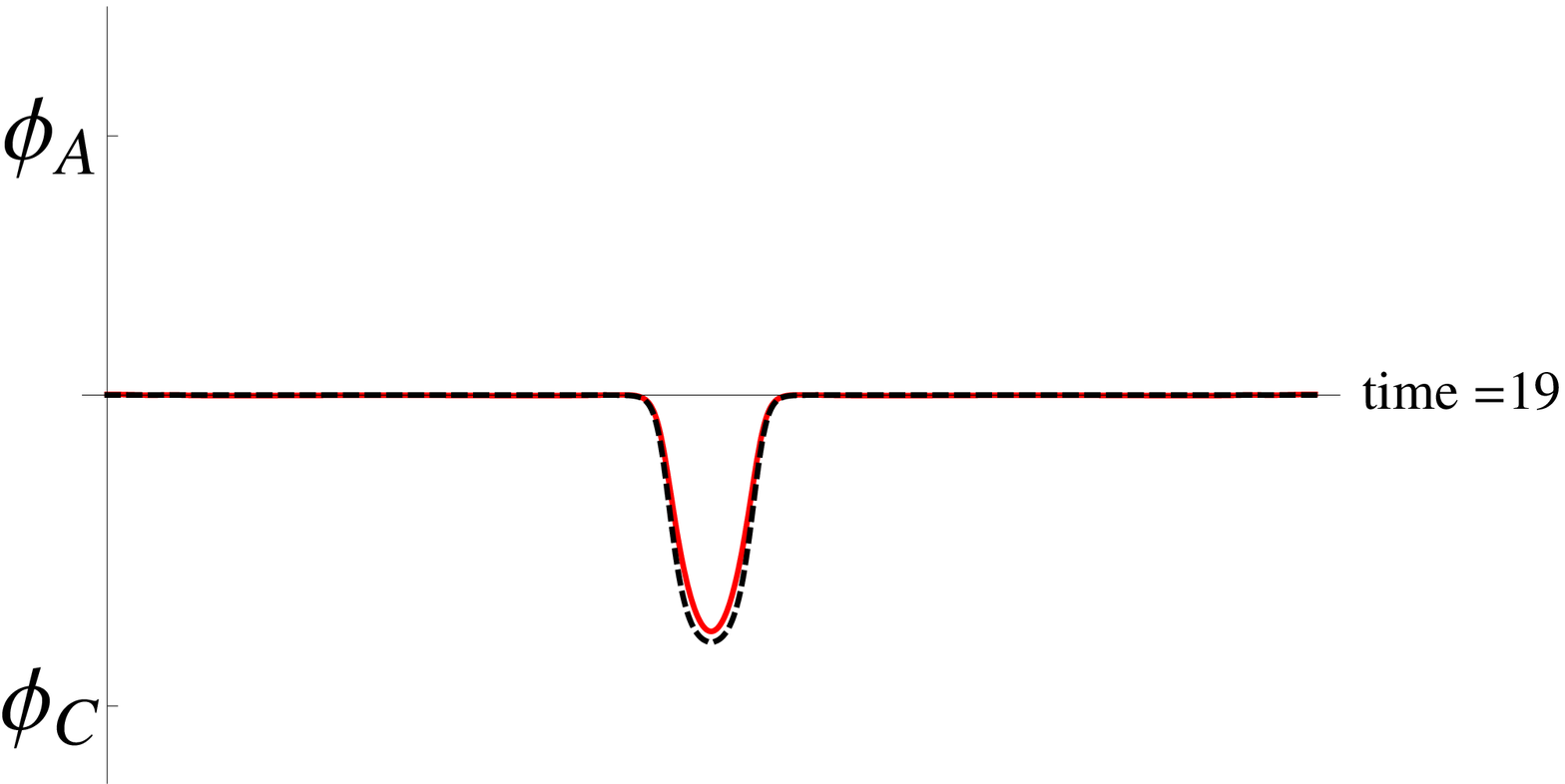}}
\subfigure[]
{\label{march6}\includegraphics[width=.25\textwidth]{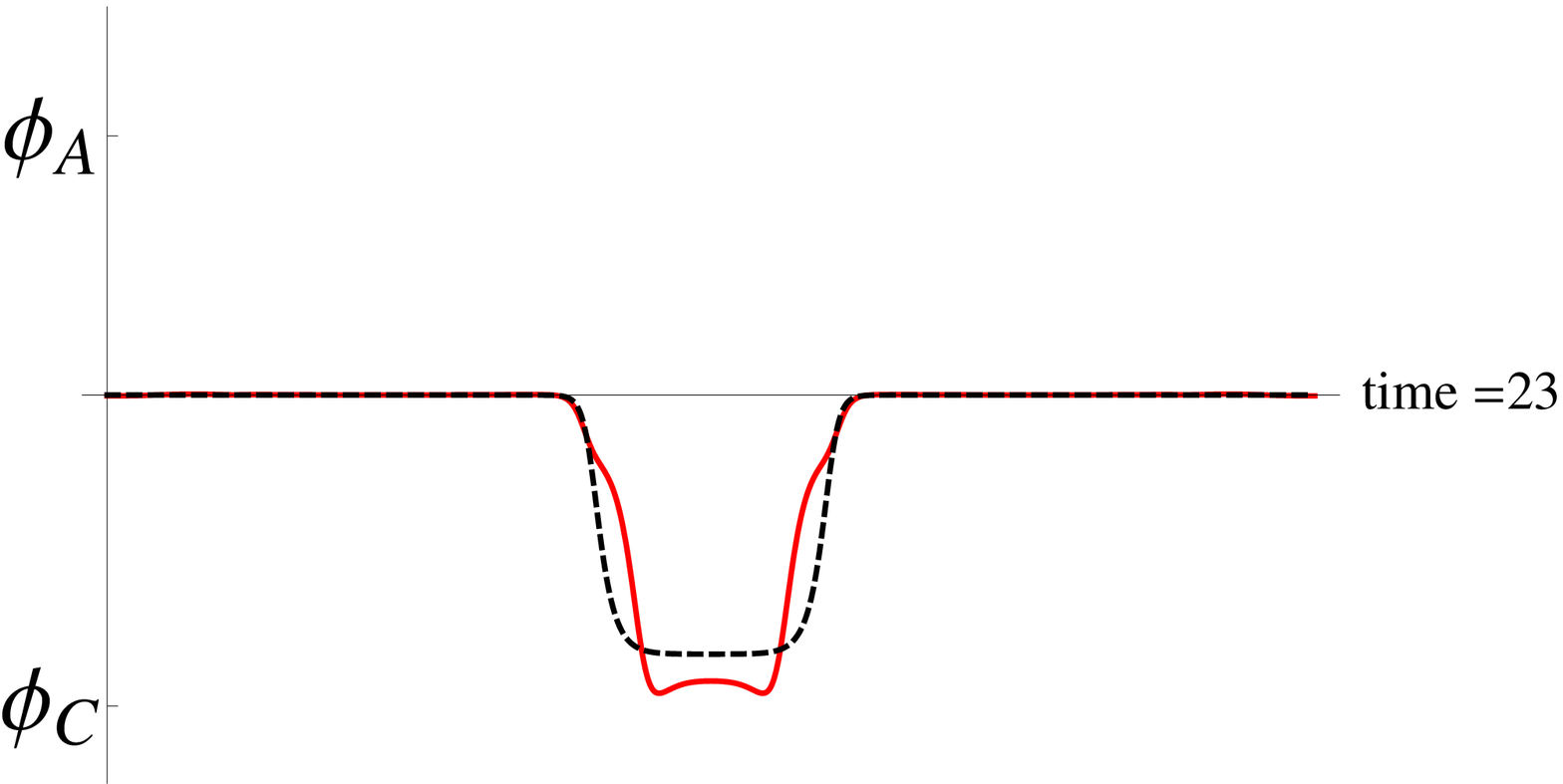}}
\caption{These are snapshots of the $1+1$ dimensional numerical solution
to the collision of a soliton and an anti-soliton,
interpolating between $\phi_B$ and $\phi_A$, as in Fig. \ref{fig:Tompot}.
They are time ordered from panel (a) to (f). The red line shows the numerical solution.
The black line shows the free passage or linear superposition prediction Eq. (\ref{phisum}).
Here, the minimum energy condition Eq. (\ref{gammacondition}) is satisfied:
the threshold for the potential chosen is $\gamma_{\rm crit} \approx 1.7$, the
$\gamma$ used in the simulation is $2.3$.
}
\label{marchmovie}
\end{figure*}

In this section, we display a number of 
$1+1$ and $3+1$ dimensional numerical solutions
to test our analytic results so far, namely
the free passage kick Eq. (\ref{1Dkick}), and the minimum
energy condition Eq. (\ref{gammacondition}). 
We adopt the following potential (Fig. \ref{fig:Tompot}):
%In this section, we show that the energy condition of the 
%previous section follows even in $3+1$-dimensional numerical simulations.  
%Consider the following model, 
\begin{equation}
\label{fulltoypot}
V_3(\phi) = \frac{\lambda}{4}\phi^2 \left(\phi-\phi_0\right)^2\left(\phi+\phi_0\eta \right)^2 + \epsilon \lambda \phi_0^5 \left(\phi-\phi_0\right),
\end{equation}
where we label the subscript $3$ to signify that there are three minima associated with this model.  There are four free parameters in this model, $\lambda$ and $\phi_0$ represent the overall scaling of the height and breath of the potential, $\epsilon$ breaks the degeneracy of our minima and provides the pressure that accelerates the bubble walls, and $\eta$ breaks the symmetry of the potential 
(making the minima {\it unevenly} spaced and the barrier heights {\it unequal}).
This is a modification of the canonical 3-minima model of EGHL  which can be recovered by setting $\eta=1$.  
Here, we adopt $\eta = 1.15$. 
We have neglected an additive constant to the potential, 
since we are working in the Minkowski limit.

\begin{figure*}[tb]
\subfigure[]
{\label{retreat1}\includegraphics[width=.25\textwidth]{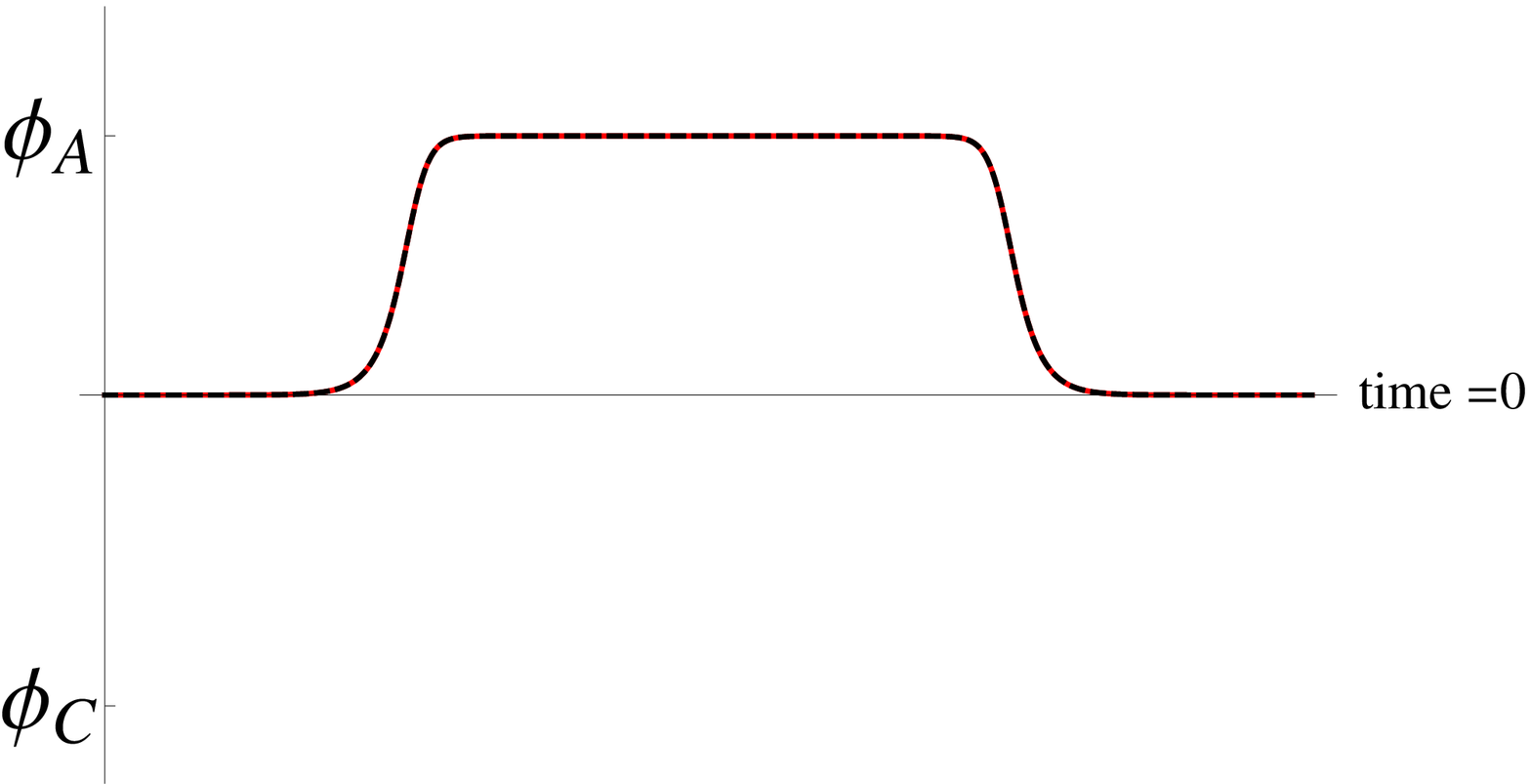}}
%\vspace{0.01in}
%
\subfigure[]
{\label{retreat2}\includegraphics[width=.25\textwidth]{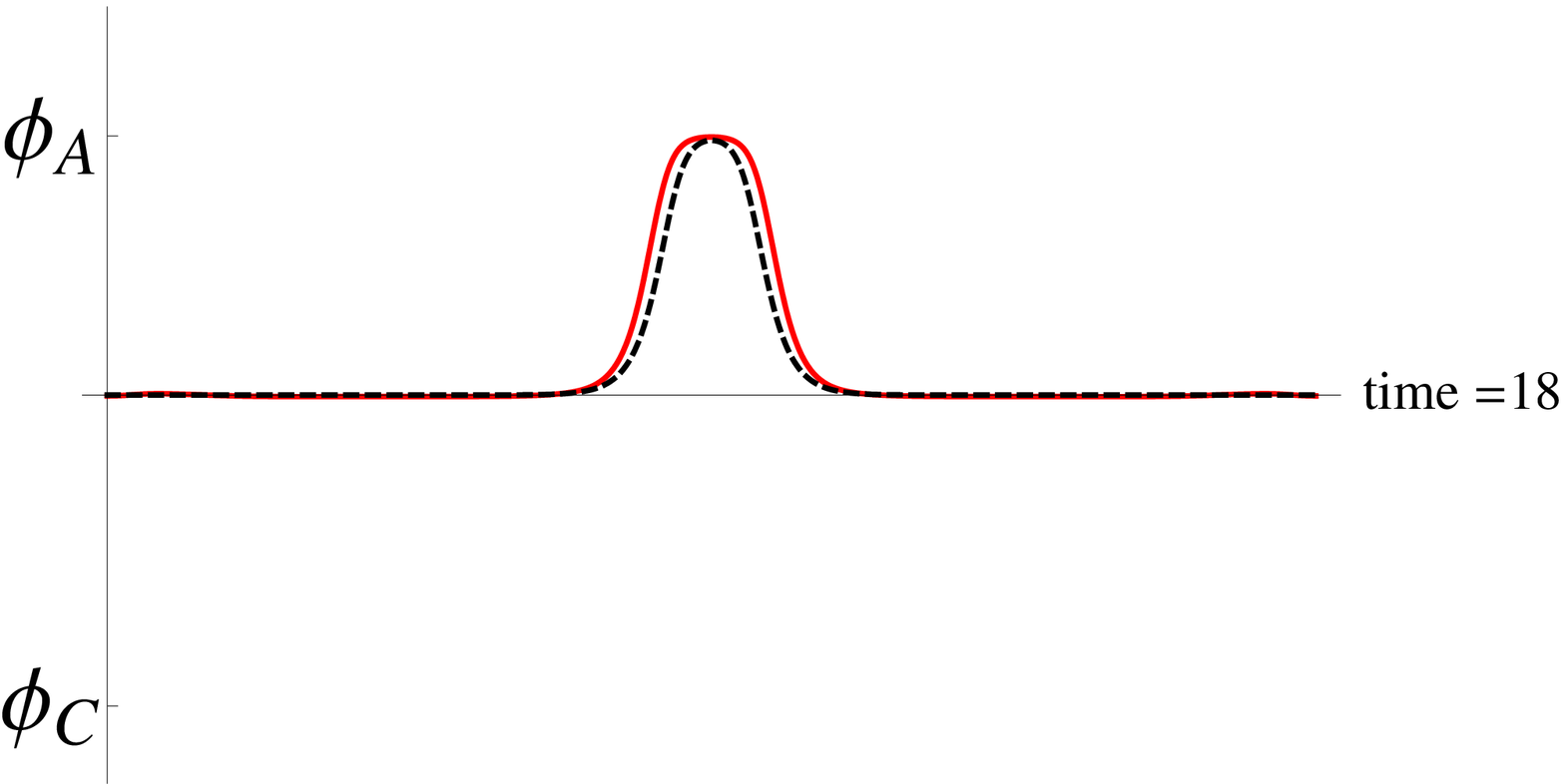}}
%\hspace{0.01in}
%
\subfigure[]
{\label{retreat3}\includegraphics[width=.25\textwidth]{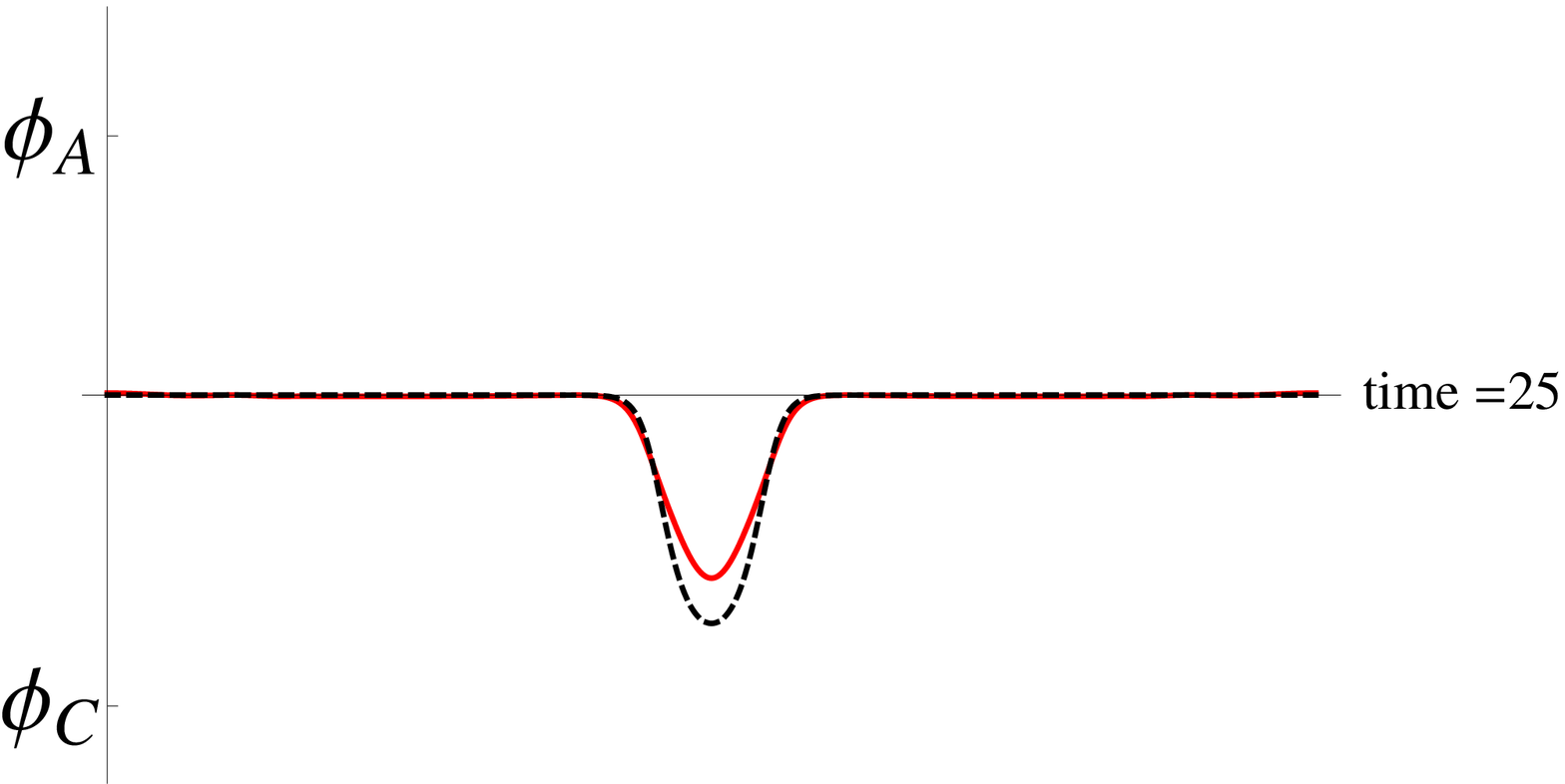}}
\subfigure[]
{\label{retreat4}\includegraphics[width=.25\textwidth]{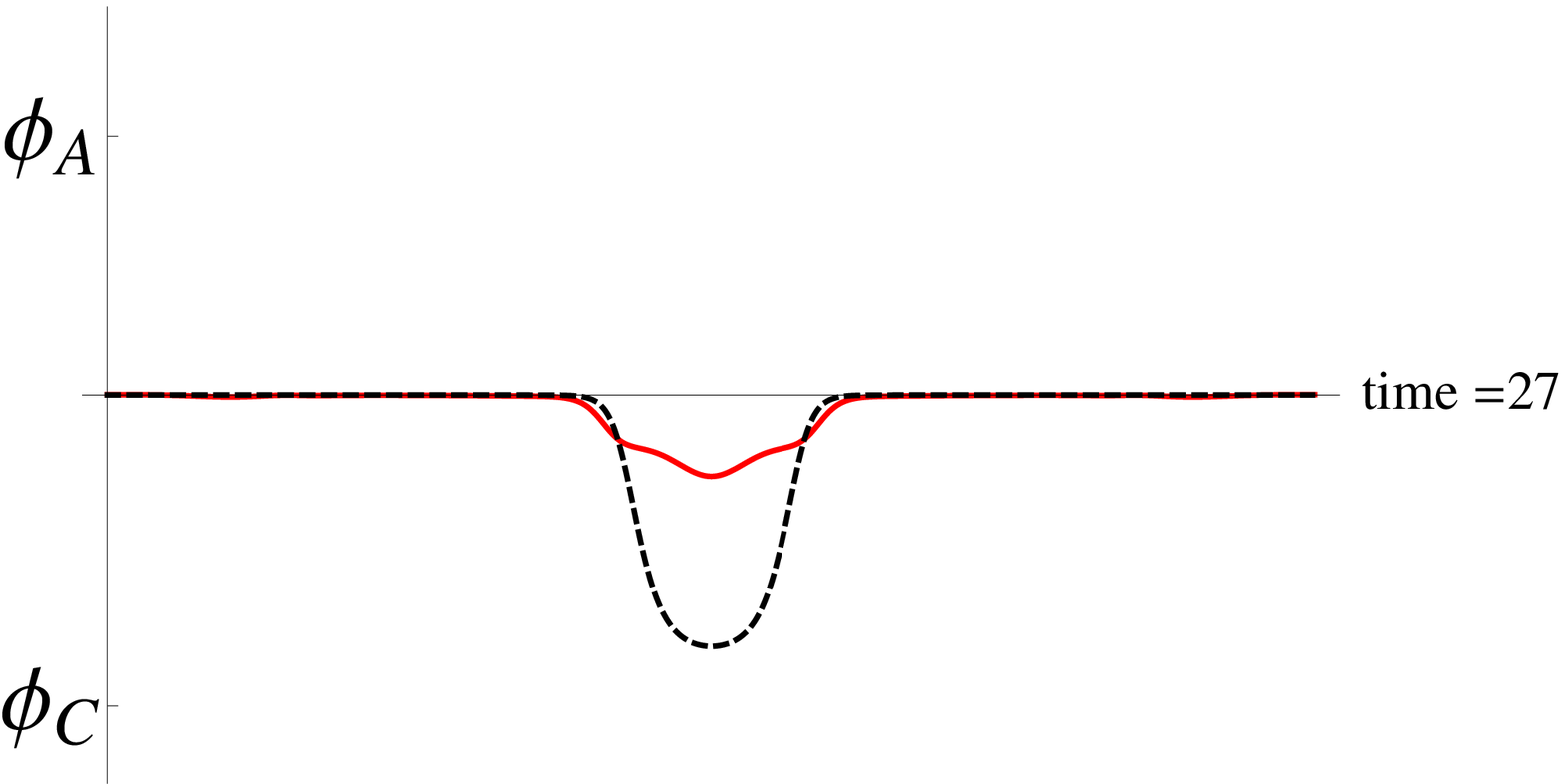}}
\subfigure[]
{\label{retreat5}\includegraphics[width=.25\textwidth]{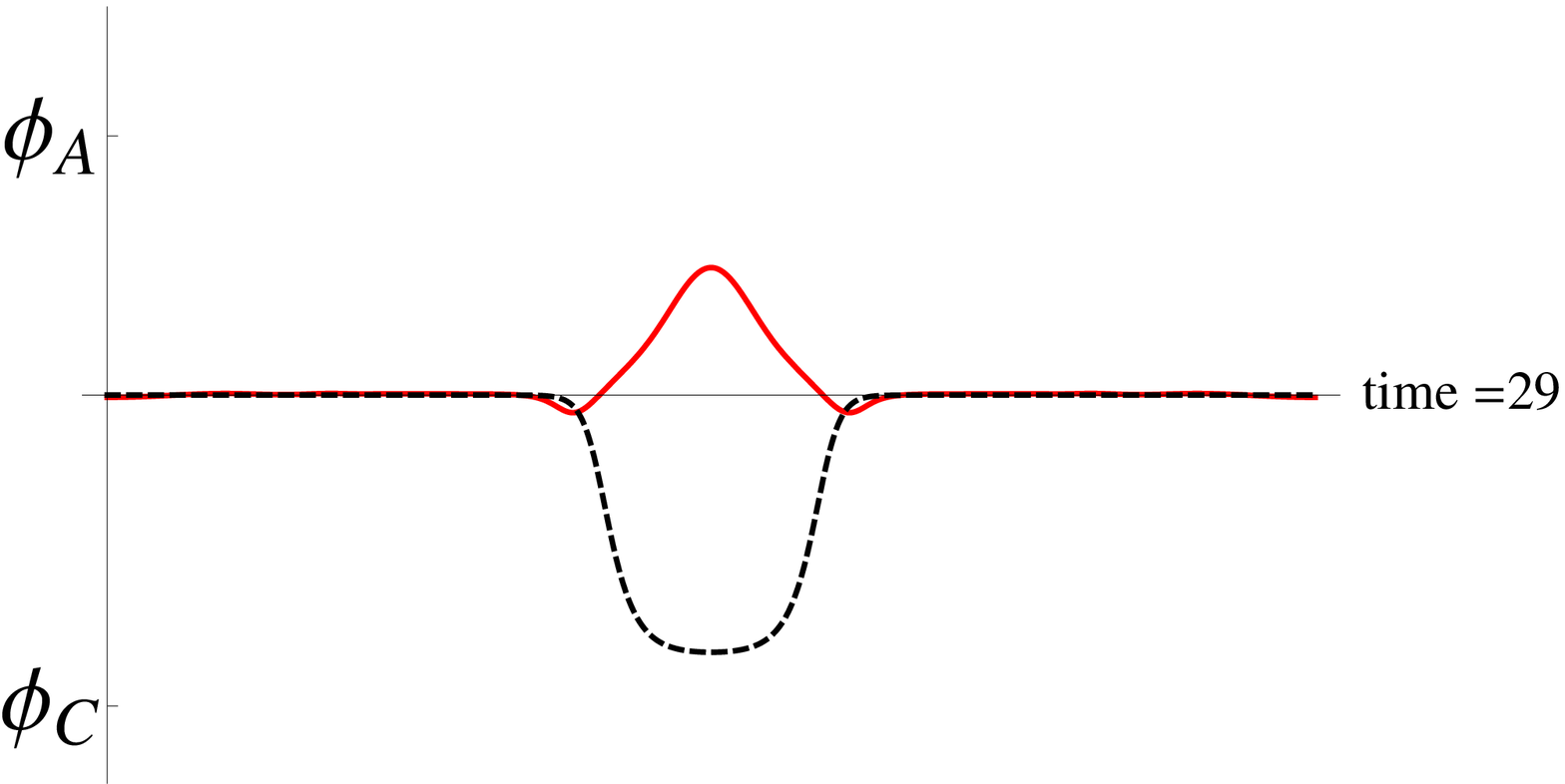}}
\subfigure[]
{\label{retreat6}\includegraphics[width=.25\textwidth]{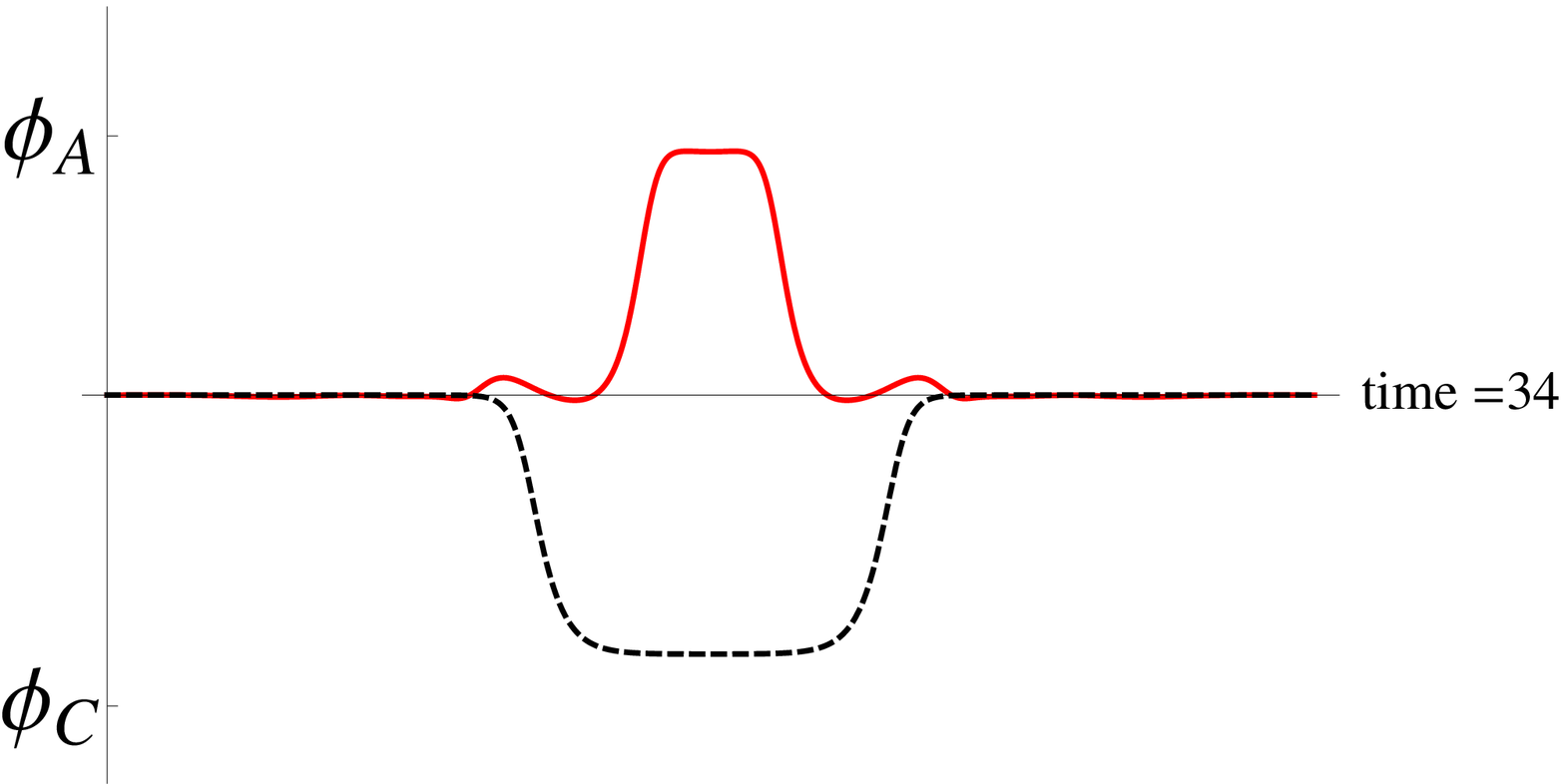}}
\caption{
Analog of Fig. \ref{marchmovie} except that the $\gamma$ used is $1.4$, and therefore
the free passage kick fails, i.e. the field retreats back to $\phi_A$ eventually.
}
\label{retreatmovie}
\end{figure*}

\begin{figure*}[tb] %  figure placement: here, top, bottom, or page
\subfigure[]{
   \centering
   \includegraphics[width= 6.8in]{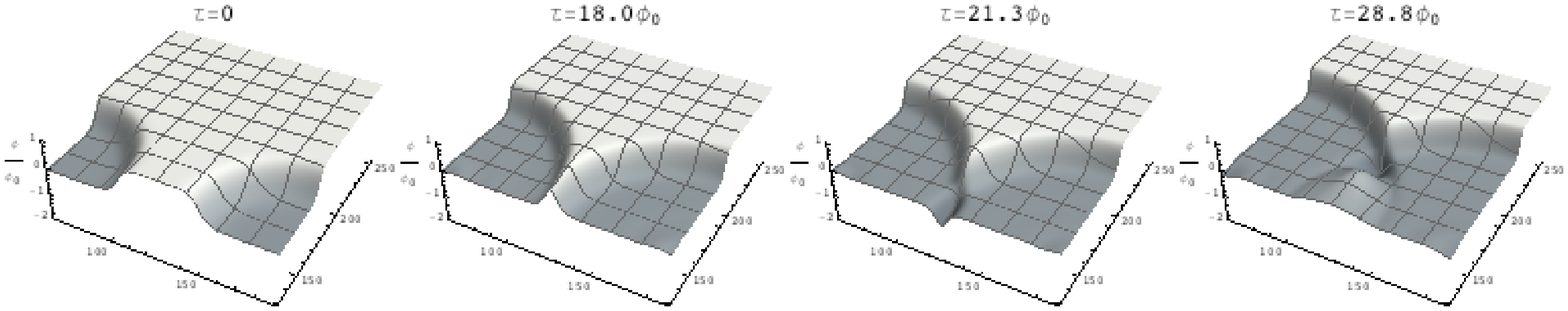} 
%   \caption{The time evolution of two bubbles whose centers are separated by $3.0R_0$ so that the bubbles achieve $\gamma = 1.5$ just before collision. Each diagram shows a 2D slice of the field values at equal time.  This uses model (\ref{fulltoypot}) with $\eta=1.15$}
   \label{fig:fpretreat}}
\subfigure[]{
   \centering
   \includegraphics[width= 6.8in]{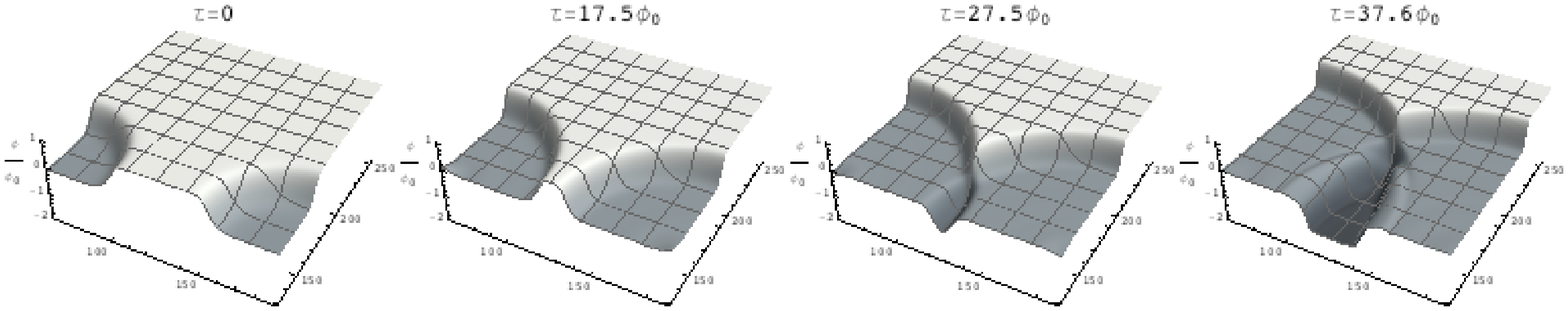} 
   \label{fig:fpmarch}}
   \caption{The top (bottom) panels show the time evolution of two bubbles whose centers are separated by $3.0R_0$ ($3.4R_0$) so that the bubbles achieve $\gamma=1.5$ ($\gamma = 1.7$) just before collision. Each diagram shows a 2D slice of the field values at equal time.   This uses model (\ref{fulltoypot}) with $\eta=1.15$}
\end{figure*}

Let us start with results of the $1+1$ dimensional simulations, for which $\epsilon$
is set to zero. Here, we collide solitons that interpolate between the appropriate minima.
The results are shown in Fig. \ref{marchmovie}
for a collision that meets the minimum energy condition.
We can see that the free passage approximation works very well,
except in the very last snapshot. This is after the free passage kick
has been successfully realized, the field in the collision region
is taken to a point just shy of the new vacuum $\phi_C$.
More precisely, the free passage kick 
takes the field to $-\phi_0$, whereas $\phi_C$ is actually at
$-\eta\phi_0$. The field subsequently rolls
down the potential to reach $\phi_C$ --- this subsequent evolution
is not well described by free passage, as expected.

An analogous set of snapshots for a subcritical collision is
shown in Fig. \ref{retreatmovie}. Here, the collision energy
is not enough for the free passage kick to succeed.
Note however that the field does make a brief excursion
in the direction of $\phi_C$, but quickly retreats, resulting
eventually in an outgoing pair of solitons that are just like
the incoming ones. The free passage approximation does not
give the correct post collision field configuration, again as expected.

For the $3+1$ bubble (as opposed to soliton) simulations,
we follow the method of EGHL for our numerics; the only difference being the potential is given by (\ref{fulltoypot}) and we neglect the expansion of the box for clarity as it is not essential to our work here -- we have checked numerically\footnote{We will be happy to provide the simulation results if the reader is interested.} that the results do not vary much with expansion turned on in the presence of an overall vacuum energy much bigger than the energy difference between the minima. We find that 
the analytic arguments and 1+1 soliton simulations are replicated very well in the 
$3+1$ dimensional simulations. The $3+1$ simulations are crucial in confirming
that there are no gross instabilities that might be missed in the $1+1$ simulations.

We nucleate two bubbles at zero wall velocity and allow the pressure difference 
($\epsilon = 1/30$) to accelerate the walls to the desired collisional $\gamma$ determined by the initial separation distance $d$.  
%Since the third minima is \emph{farther} in field space, we see that free passage does not assure that field values make it to $\phi=-\eta\phi_0$, the furthest it can go after free passage is to $-\phi_0$. This is far enough for it to go over the peak of the second barrier. Hence, as we discussed in the previous section, the completion of the transition relies on the collision energy of the walls.  
Figures \ref{fig:fpretreat} and \ref{fig:fpmarch} show the  time lapse of the results of two simulations with $\gamma =1.5$ and $\gamma=1.7$, with a critical $\gamma_{\mathrm{crit}} \approx 1.6$
(slightly different from the value in $1+1$ because of the difference in $\epsilon$). 
Comparing this to the theoretical expectation Eq. (\ref{gammacondition}) fixes
$\alpha^{-1}\approx 5.5$. 
%To numerically test the viability of transition to $\phi=-\eta \phi_0$ under any circumstances, we bump up the velocity of the walls at the time of collision.  

In the first simulation, we can clearly see the field attempts to
execute the free passage excursion, but due to the low collision velocity, the field does not fully transition and retreats back into the middle vacuum $\phi_B$.
This final behavior is 
unlike the soliton case, where the field retreats all the way back to $\phi_A$ ---
the difference arises because the soliton simulations have degenerate vacua, whereas
$\phi_A$ has the highest vacuum energy in the bubble simulations.
In the second simulation, the bubbles are nucleated at $d=3.4R_0$ resulting in a  $\gamma=1.7$ at the time of collision. In this case, there is sufficient kinetic energy in the fields during the collision to ensure a transition to the lowest minimum.

\section{Production of Multiple Walls/Solitons}
\label{multiple}

\begin{figure}[tb]\centerline{\epsfxsize=9cm\epsffile{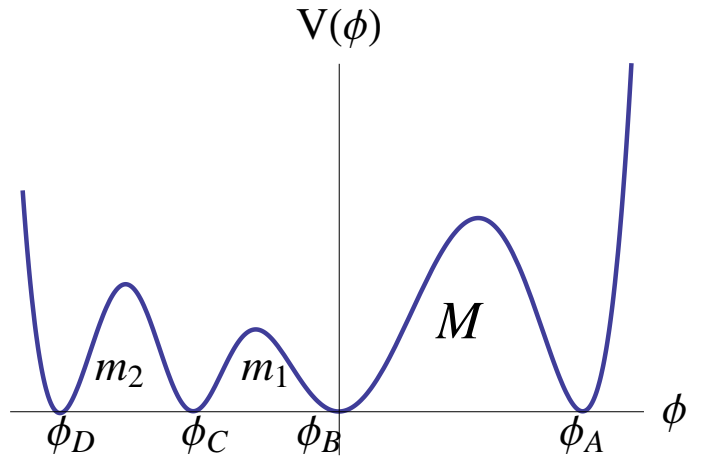}}
\caption{A potential with four minima at $\phi_A$,$\phi_B$,$\phi_C$ and $\phi_D$. Here, 
$M$, $m_1$ and $m_2$ label the rest mass of the corresponding solitons.}
\label{fig:splitpot}
\end{figure}

Suppose that in the course of the $\delta \phi$ excursion, the field passes over more than one minima (and hence more than one barrier, see Fig. \ref{fig:splitpot}), then given a sufficiently energetic collision, multiple barriers can be overcome and the end result is the production of multiple solitons each moving at a different velocity.  Much of the discussion in the previous sections applies : the field is in free passage in the initial moments after the collision, and as the approximation breaks down, begins to feel the potential and evolve accordingly. However, the field now traverses over more than one barrier, and if the energy of the collision is high enough such that the field ends up transitioning over more than one barrier, then the free-passage ``wall'' may split into two (or more) solitons in the aftermath.

Nevertheless, this does not mean that these solitons will be long-lived: the splitting of the free passage wall into two or more solitons must obey both energy \emph{and} momentum conservation, and hence may result in the soliton with field values furthest away form the original soliton to move backwards towards its mirror counterpart and annihilate each other. In the case of the solitons splitting into two, this effect can be understood by considering the solitons as massive particles and the splitting as the decay of a massive particle (the original free passage soliton) into two less massive particles. This is then simply a kinetic problem which we can solve as follows.

Consider a potential (Fig.\ref{fig:splitpot})  where the incoming soliton rest mass is $M$ and the two outgoing solitons have rest masses $m_1$ and $m_2$. According to the free passage approximation, the incoming soliton passes through the potential barriers maintaining its field profile and velocity. Since the field now traverses a different part of the potential, this profile is no longer a solitonic solution, thus its effective rest mass is in principle undefined. To make progress, we assume that this ``soliton'' instantaneously\footnote{This assumption is 
in keeping with the spirit of the free passage approximation, although one can argue that the decay is a 2-step process in the following sense -- the incoming solitons pass through each other in free passage, and then either gain or lose mass due to the fact that their profiles no longer traverse the original potential resulting in a change in velocities, before decaying.} decay in the \emph{center of mass} frame (denoted by primes) of the soliton with its incoming 4-momentum
\begin{equation}
P^{\mu} = (M,0,0,0).
\end{equation}
Meanwhile, the two outgoing solitons possess the following 4-momenta
\begin{equation}
P^{\mu}_1 = (\gamma_1' m_1, \gamma_1' m_1 u_1',0,0)~,~P^{\mu}_2 = (\gamma_2' m_2, \gamma_2' m_2 u_2',0,0).
\end{equation}
Conservation of energy and momentum $P^{\mu} = P^{\mu}_1+P^{\mu}_2$ then allows us to solve for $u_1'$ and $u_2'$, which we can then transform back into the \emph{center of collision} frame\footnote{In this example, we have assumed that both incoming solitons have identical rest masses, hence this frame 
(where the incoming velocities are equal and opposite) is also the center of mass frame for the total system.}. Assuming then the first soliton $m_1$ is formed with velocity moving away (defined to be positive) from the center of collision frame, then the second soliton has the following velocity 
\begin{equation}
u_2 = \frac{u_2'+u}{1+u_2'u} \label{eqn:2solitoncon}
\end{equation}
which must be $> 0$ for the second soliton-antisoliton pair (mass $m_2$) 
not to self-annihilate. 

To be specific, consider the following modification to our toy potential Eqn. (\ref{fulltoypot})
\begin{equation}
\label{fourminima}
V_4 (\phi) = V_3(\phi) - \delta \exp\left(-\frac{(\phi-\phi_a)^2}{b^2}\right),
\end{equation}
where $\phi_a$ defines the location of an additional metastable minimum, and $b^2$ defines the width of that minimum.  Although $\delta$ looks like a free parameter in this model, we fix it so that when $\epsilon=0$ our minima are all degenerate, hence,
\begin{equation}
\delta =  \frac{\lambda}{4}\phi_a^2 \left(\phi_a-\phi_0\right)^2\left(\phi_a-\phi_0\eta \right)^2.
\end{equation}

\begin{figure*}[tb]
\subfigure[]
{\label{fig:splitbefore}\includegraphics[width=.45\textwidth]{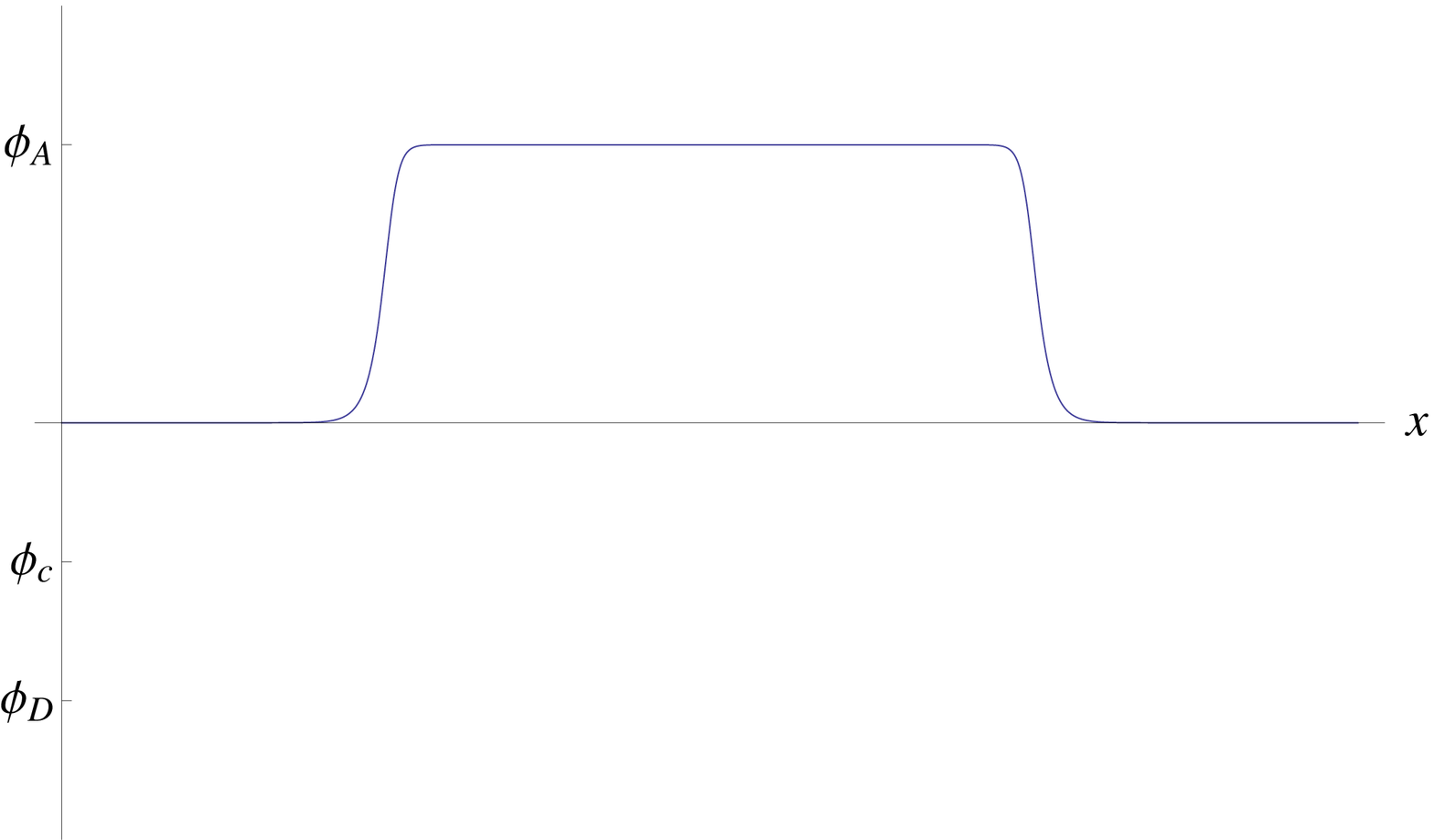}}
\subfigure[]
{\label{fig:splitafter}\includegraphics[width=.45\textwidth]{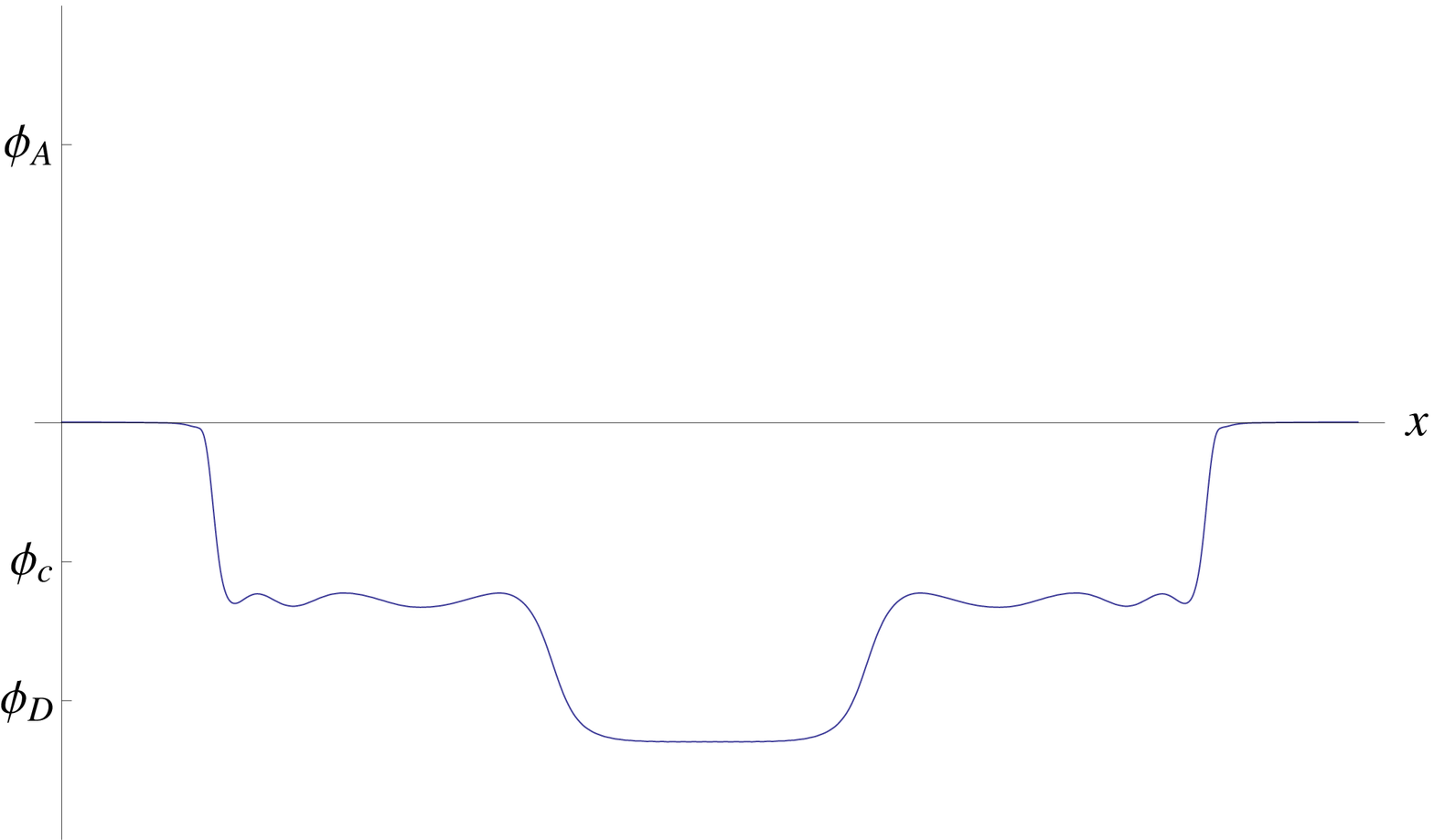}}

\caption{The interaction of a soliton-antisoliton pair in the presence of the potential denoted in Fig. \ref{fig:splitpot} and Eq. (\ref{fourminima}) with $\eta=1.15$, $\phi_a=0.65\phi_0$, $b^2=0.8\phi_0^2$.  The incoming soliton-antisoliton pair spans the barrier from $\phi_{A}$ to $\phi_{B}$ (left figure).  Given a sufficiently energetic collision -- $\gamma>2.3$ for this potential -- the field can transition to $\phi_D$, forming a pair of solitons (and a pair of
anti-solitons) in the process  (right figure). In general, the outgoing pair of solitons possess different velocities depending on their rest masses and 
the actual collision energy. For some potentials, the process can be thought of as annihilation of the two incoming solitons to form a pair of temporary free passage solitons with the same mass, and then the decay of these solitons into two pairs of less massive solitons. Since the interaction is not completely elastic, scalar radiation visible as superimposed small perturbations is emitted. }
\label{fig:splitinter}
\end{figure*}

This potential is equivalent to the soliton potential shown in Fig. \ref{fig:splitpot}. For a model with $\eta=1.15$, $\phi_a=0.65\phi_0$ and $b^2=0.8\phi_0^2$, one can calculate that for $u_2=0$ (i.e. the critical splitting velocity) $\gamma=2.3$. Colliding the solitons at $\gamma=2.6$, we can use Eq. (\ref{eqn:2solitoncon}) to find that the second soliton will have a velocity of $u_2\approx 0.2$ and hence a splitting will occur, a result which is numerically confirmed in Fig. \ref{fig:splitinter}.

\begin{figure*}[tb] %  figure placement: here, top, bottom, or page
   \centering
   \includegraphics[width= 6.8in]{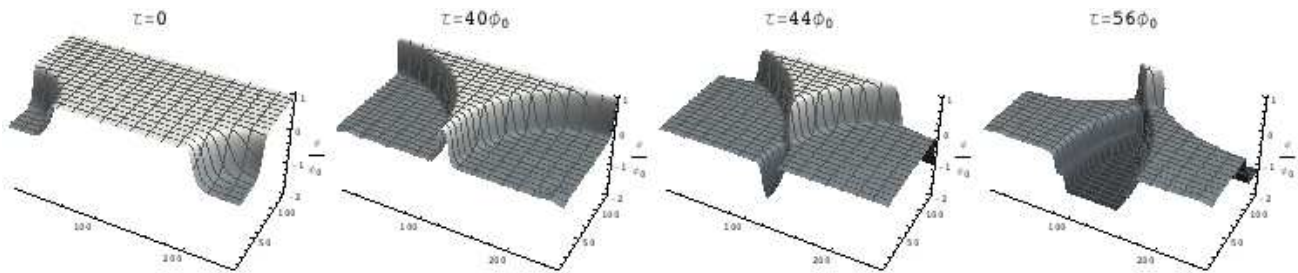} 
   \caption{The time evolution of two bubbles whose centers are separated by $5.2R_0$ so that the bubbles achieve $\gamma = 2.6$ just before collision.  This uses the model (\ref{fourminima}) with $\eta=1.15$, $\phi_a = 0.65 \phi_0$, $b^2=0.8\phi_0^2$.  Periodic boundary conditions produce multiple images of this collision in these panels.}
   \label{fig:multiplesolitons}
\end{figure*}

%For the cases of splitting into more than two solitons,
%or the splitting into two solitons with rest masses greater than 
%the incoming soliton $m_1+m_2 > M$, 
%there is no simple way to predict the outcome -- one has to resort 
%to numerical simulations in these cases.

As a final numerical test, we show that such splittings occur even in $3+1$ dimensional simulations.  We use the same potential (\ref{fourminima}) as in the solitonic case, except that we have added in a small linear tilt term $(1/30)\lambda(\phi-\phi_A)\phi_0^5$.  Since we anticipate that a highly relativistic collision is necessary to produce multiple domain walls, we begin by nucleating two bubbles at a separation of $d=5.2R_0$.  Figure \ref{fig:multiplesolitons} shows the time evolution of these events.  We can clearly see the free passage of the field during the collision, and then the creation and acceleration of two domain walls, making a double bubble.

Finally, we comment on an interesting possibility. Consider the case such that the collision $\gamma$ is barely insufficient to satisfy Eq. (\ref{eqn:2solitoncon}). In this case, a fully formed soliton-antisoliton pair between $\phi_C$ and $\phi_D$ are formed but self-annihilate after at while. This secondary collision can be treated as any other collision -- if the right conditions are satisfied then we can form a pocket of $\phi_B$ vacuum, creating a ``bubble within a bubble'' scenario. 

%Numerically, the answer is yes -- the final result of such a multi barrier transition is the production of multiple solitons (FIGURE will be useful), each moving at a different velocity. Much of the discussion in the previous section naturally 

%Sometimes, the outgoing pair of walls predicted by free passage
%are unstable towards splitting into multiple ones, as is the case
%if there are additional local minima in the course of the excursion
%$\delta\phi$.
%{\bf Eugene: please add discussion on splitting off of solitons...}

\section{DISCUSSION} \label{sect:discussion}

To briefly summarize, we find that linear superposition of wall profiles
provides a good approximation to what happens at a high speed collision.
The free passage of wall profiles implies a field excursion in the
collision region that is given by 
Eq. (\ref{multidimkick}).
For this free passage kick to be successfully
realized, the collision must exceed a certain minimum energy threshold
given by Eq. (\ref{gammacondition}). 
Beyond this threshold, the size of the kick itself is
independent of the collision speed, i.e. one cannot obtain arbitrarily
large field excursions by pumping up the collision energy.
When the minimum energy condition is met, the free passage kick can cause
a transition into a new vacuum, if the kick takes the scalar field to
within the new vacuum's basin of attraction.
Interesting possibilities arise if the excursion traverses over
several new vacua, and they can be understood by similar arguments
(\S \ref{multiple}). 
We have verified these statements
using numerical computations of soliton collisions in $1+1$ and bubble
collisions in $3+1$. 

These findings have many interesting theoretical and observational implications. 
Let us discuss some of them. 

\subsection{Including Gravity} \label{sect:gravity}

Our simulations used a flat, non-expanding background.  To study a classical 
transition in the cosmological context, first we need to know the necessary
modifications once we include gravitational effects.

Since the collision happens on a short time scale, the corrections from 
gravity will not be in the 
collision itself, but in the domain wall motions before 
and after the collision.  Without gravity, it appears that
arbitrarily high barriers can always be traversed via collisions,
as long as the incoming boost $\gamma$ is large enough.
In an expanding background, however, 
there is a maximum boost for every collision,
\begin{equation}
\gamma_{\rm Max} \sim (R_0 H_A)^{-1}~,
\end{equation}
where $R_0$ is the initial size of the bubble B, and $H_A$ is the hubble 
expansion rate for the parent vacuum A.  Therefore, it is easy to generalize 
our results to an expanding background, as long as we maintain an additional 
condition : the critical $\gamma$ to make the transition in flat space has to 
be smaller than $\gamma_{\rm Max}$ in the expanding background, otherwise there
will be no transition.

After the transition, there are more complications.  For example, if 
$V_C>V_B$, the domain walls can turn around and may or may not collide with 
each other 
again, thereby sealing the vacuum C region.  Also when the domain wall is 
heavy, $m_{BC} < |V_C-V_B|$, gravity allows it to accelerate and eventually 
run away from both sides.  These behaviors are hard to keep track of in 
lattice simulations.  An analytical study on these questions can be found in  \cite{JohYan10}.

\subsection{Cosmology After a Collision} \label{sect:cosmology}

Many recent studies discussed the possibility of observing signals from 
bubble collisions in our past.  Several cases are summarized in 
\cite{AguJoh09}.  One important issue is 
whether slowroll inflation can be maintained
after a violent disturbance 
like a bubble collision.  The scalar field behavior we 
study here provides useful intuition to address such a problem.

If the field range of slowroll on the potential is super-planckian, which is
usually known as a large-field slowroll inflation, then it is quite stable 
against disturbances as we do not expect the vacuum separation 
(which determines the collision induced kick) to be 
super-planckian. A sufficiently large flat region in the potential 
post-collision would provide a buffer for slowroll to continue unscathed. 
On the other hand, if the slowroll
range is smaller, then it is more delicate.  One example in 
\cite{AguJoh09} seems to suggest that a collision always ends small-field 
slowroll inflation.  Here we would like to argue the opposite---\emph{Having 
small-field slowroll inflation after a collision is as plausible as 
having it after a Coleman-De-Luccia tunneling.}

We start from Figure 7 in \cite{AguJoh09}, which the authors interpreted as 
the collision pushing the field from slowroll directly to the end of
inflation.   
%I-Sheng, I am afraid the subtlety of these different interpretations
%will be lost on the reader, unless the reader actually open up their paper
%and read it side by side - an unlikely event. So I comment this out.
%
%Looking at their potential (Figure 4 in the same paper), the 
%interpretation implies that the field goes in the opposite direction of a free 
%passage.  This is not a contradiction -- but our physical interpretation 
%here will be different.  For the  authors in \cite{AguJoh09} 
%the ``collision region'' 
%refers to the future lightcone of the collision, while in 
%this paper the same term 
%refers to the region generated by domain walls just crossed each other.
The free passage tells us that right after the domain walls cross each other,
the field ends up within a region (possibly very) near the parent vacuum (see Figure 7 of \cite{AguJoh09}). Because of the pressure difference, such a region 
undergoes oscillations and re-
collisions as in \cite{HawMos82a}. In the figure cited above, the first 
two collisions are visible but the rest of them are too small to be resolved, 
forming an effective domain wall\footnote{We have reproduced this result in both 3+1 and 1+1.}.  The pressure difference 
between the bubbles on left and right accelerates this effective domain wall 
to the right.  
%Everything obeys the free passage and general laws of domain 
%wall motions.  

The creation of this domain wall is highly dissipative -- scalar radiation is emitted from the point of initial collision and subsequent re-collisions, and this scalar radiation is large enough in amplitude to disrupt the small field inflation, causing the inflaton to topple off its delicately balanced potential and thermalize.  In other words,  the failure of small-field slowroll inflation appears to be a result of the 
disturbances of the repeated re-collisions hidden in the effective domain wall. 

\begin{figure}[tb]\centerline{\epsfxsize=8cm\epsffile{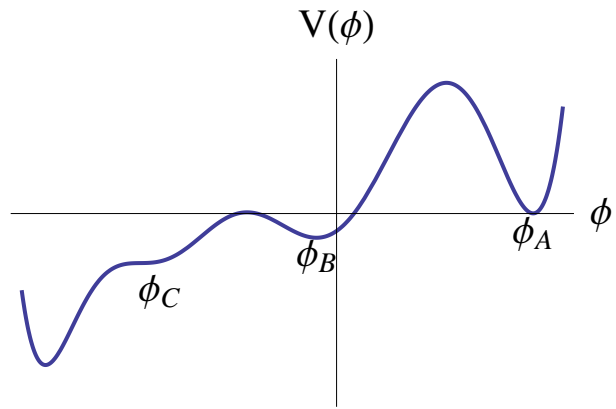}}
\caption{A potential with vacua $A,B$ and a small field slowroll potential 
around $C$.}
\label{fig:ABCroll}
\end{figure}

%One can imagine setting up a perfect collision where the domain wall collisions is super efficient, and hence very little radiation is leaked. 
Nevertheless, as we have  shown in this paper, counter to the intuition 
that collisions are 
violent, classical transitions can be exceptionally gentle
(recall the fairly homogeneous post-collision region in our
$1+1$ and $3+1$ simulations). This gentleness 
might live in harmony with the delicate small-field potential.
Consider a potential in Fig.~\ref{fig:ABCroll}.  If we arrange that
$\phi_A-\phi_B=\phi_B-\phi_C$, then by free passage, a collision between
two bubbles of $B$ naturally starts slowroll inflation in $\phi_C$.
One might argue that such a setup is finely tuned.  Let us ask a different 
question: Can a Coleman-De-Luccia tunneling from 
vacuum $B$ start slowroll inflation in 
$\phi_C$?  The answer is yes, if we arrange the potential correctly -- it is 
also finely tuned!

Although there seems to be an additional parameter $\gamma$ involved in 
collisions, we remind ourselves the following two facts.
\begin{itemize}
 \item Our simulation shows that once the transition is allowed, increasing
 $\gamma$ further has no impact on the field value right after the
 transition.  Namely, the initial condition for starting the slowroll is 
 not sensitive to the incoming boost.
 \item In a multiverse we have an infinite number of 
 collisions with different $\gamma$'s, so 
 it is naturally scanned and does not require further fine-tuning.
\end{itemize}
Therefore, all the fine-tuning is on the potential itself.  We conclude that 
starting slowroll through collisions is not more fine-tuned than starting 
it through a single tunneling.\footnote{These fine-tunings are only necessary 
in small field inflation models.  In large field models it will be equally
generic to start inflation through either tunneling or collision.}

Let us now turn to the question of the spacetime geometry inside a collisionally 
formed bubble. 
The motivation is that we might perhaps live in it, i.e. a collision
event might have constituted our beginning, our big bang.
Assuming that there is slowroll inflation inside the bubble, 
we can approximately describe the geometry as pure de~Sitter, whose 
general metric is given in \cite{FreHor07}:
\begin{equation}
ds^2 = -\frac{dz^2}{f(z)}+f(z)dx^2+z^2dH_2^2~,
\end{equation}
where 
\begin{equation}
f(z) = 1-\frac{M}{z}+H^2z^2~.
\end{equation}

When the domain walls are highly boosted, the conservation of stress-tensor 
gives us\footnote{This is exactly true for massless(null) domain 
walls \cite{LanMae01}, and  approximately true for light or highly boosted domain 
walls.}
\begin{equation}
f_A(z_c)f_C(z_c) = [f_B(z_c)]^2~,
\end{equation}
where $z_c$ is the collision radius of the two-dimensional
hyperbolic section $H_2$.  The $A, B, C$ labels
follow the labeling of vacua in the rest of this paper.
Neglecting radiation loss 
(which seems to be a good approximation from our simulations),
$M=0$ in both $f_A$ and $f_B$.  As arranged 
in the potential $V_A>V_B>V_C$, $M$ in $f_C$ is
also small and the $M/z$ term soon becomes unimportant.  The metric in region 
C is thus approximately
\begin{equation}
ds^2 = -\frac{dz^2}{1+H_C^2z^2} + (1+H_C^2z^2)dx^2 + z^2dH_2^2~.
\end{equation}  

When there is slowroll inflation in region C, this will not be the proper 
slicing to describe the universe.  Region C is the forward lightcone 
of an $H_2$ with radius $z_c$.  Following the usual prescription --- in
the case of a single bubble ---
for deriving an open de~Sitter slicing covering the future lightcone
of a point in the global de Sitter space, 
we can similarly derive the ``open'' slicing of 
the universe in region C:
\begin{eqnarray}
ds^2 &=& -dt^2 + \frac{\sinh^2H_Ct}{H_C^2}d\xi^2 \\ \nonumber
&+& \bigg(z_c \cosh H_C t + \frac{\sinh H_Ct}{H_C}
\sqrt{1+H_C^2z_c^2}\cosh\xi\bigg)^2dH_2^2
\end{eqnarray}

We can see that the metric preserves only the symmetry of $H_2$, and is both 
anisotropic and inhomogeneous.  The interesting trait is that both its 
anisotropy and inhomogeneity are correlated with its negative spatial 
curvature.
Since the universe is roughly isotropic and homogenous to 1 part to $10^5$ \cite{WMAP7}, if we are to live in such a bubble, there must exist a sufficient amount of inflation in this new universe after its birth through a collision \cite{Wal83}. While the possibility of statistical anisotropy have been discussed, at least at the level of the initial perturbations \cite{AckCar07}, our model predicts an interesting novel anisotropy in the form of an anisotropic 
spatial curvature, i.e. there exist two different curvature scales in different directions, an intriguing possibility we plan to explore in future work.
Note that an anisotropic curvature shifts the Doppler peaks in the microwave
background on all angular scales, making the effect easier to observe
than most other large scale anomalies, as recently 
emphasized by \cite{GraHar10,BlaSal10} 
in the context of other models.
%Given some slow roll inflation in the new universe, 

%It is a known fact that typical anisotropy redshifts very fast therefore hard 
%to observe today given sufficient inflation to make observed flatness \cite{WMAP7}. On the other hand, there are still chances to observe a
%non-zero curvature.  An ``anisotropy in curvature'', as we have here, might 
%also be observable. 

\subsection{Additional Ingredients to the Field Dynamics} \label{sect:addition}

There are several possible complications to the field dynamics
beyond what we have discussed.
First of all, there can be derivative interactions.
Our analysis should hold as long as the collision process
does not probe energy scales above the cut-off for these
derivative interactions, say $\Lambda$. 
This means the boosted thickness of our bubble walls
$1/(\gamma\mu)$, where $1/\mu$ is the rest-frame thickness,
must be larger than $1/\Lambda$. This implies an {\it upper limit}
to the collision speed that we can consider $\gamma \lsim \Lambda/\mu$.
When derivative interactions are important,
linear superposition such as we have used in Eq. (\ref{phisum})
no longer works. 

Another possibility is that the field metric (in the multi-field
case) is non-trivial, and therefore the excursion trajectory is
modified. A third possibility is that some other field is
coupled to our bubble scalar field, and the coupling is
such that this other field becomes massless in the course
of an excursion. Massless particles get produced\footnote{In\cite{ZhaPia10} it is treated as a direct generalization of preheating, but we remain conservative about their results.}, and could significantly modify the excursion trajectory. We plan to investigate these intriguing possibilities in the future.

\subsection{Scanning of the Landscape}\label{sect:landscape}

The collision induced excursion
offers a new mechanism for scanning a landscape of many vacua.
Our results here touch on two aspects of this scanning.
One is that even very high barriers can be overcome by a collision
as long as the collision is relativistic enough (but subject
to constraints from expansion, and from derivative interactions).
It could well be that certain vacua surrounded by high barriers
are more likely to be populated by collisions rather than direct
tunneling. A second important feature of this 
scanning is that it is bounded --- the collision induced excursion
does not become arbitrarily large by raising the collision speed.
Rather, it is controlled by a simple vector sum rule (Fig. \ref{multiscalarkick}),
which tells us that the excursion can only be as large as the
field difference between the parent and bubble vacua.
In theories in which tunneling between vastly separated vacua is
common \cite{BroDah10}, large collision excursions are also possible.
Finally, this excursion has a specific direction, and it is interesting to ask what is the typical basin of attract there.  In\cite{AguJohLar09} is it  suggested that due to the universal dilatonic runaway direction in string theory inspired models, classical transition can lead to decompactification of extra dimensions.

\subsection{Multiple Collisions} \label{sect:multi}

\begin{figure}[tb]
\centerline{\epsfxsize=6cm\epsffile{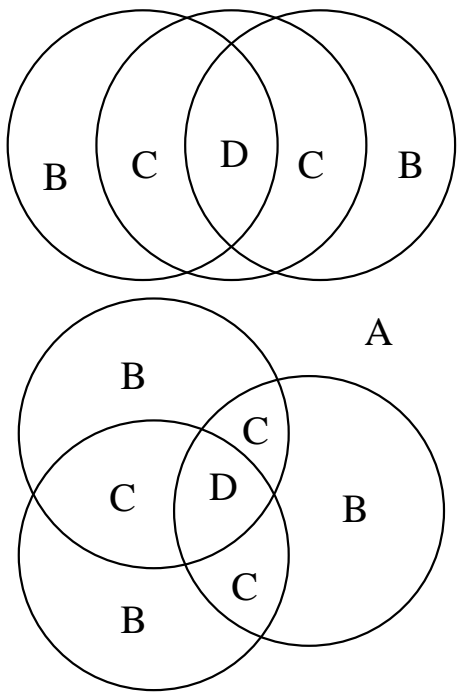}}
\caption{A spacelike slice through multiple, overlapping collisions.  
Following the vector sum rule, we have:\\
$\vec \phi_C=2\vec \phi_B-\vec \phi_A$, $\vec \phi_D=3\vec \phi_B-2\vec \phi_A$.}
\label{figure:CFT}
\end{figure}

The free passage approximation and the resulting vector sum rule in 
section \ref{freepassage} is a simple fact of the field dynamics.  It applies to
any solitonic objects interpolating between local minima.  We can 
arrange a potential in which a collision is followed by further collisions,
leading to a nested set of classical transitions.
A spacetime 
structure with multiple, overlapping collisions follows simply from the vector
sum rule, as shown in Figure~\ref{figure:CFT}.

A particularly intriguing version of this occurs when a bubble is nucleated in
the presence of a compact dimension \cite{Bro08}. The bubble grows, wraps
around the compact dimension, and eventually collides with itself. A 
new vacuum opens up 
between the outgoing pair of walls. These walls eventually collide 
after traveling
through the compact dimension. Further transitions and collisions follow,
as far as the potential allows. This provides a novel way to realize an old
idea by Abbott \cite{Abb85}, by classical transitions rather than tunneling.

\begin{figure*}[tb]
\epsfxsize=5cm\epsffile{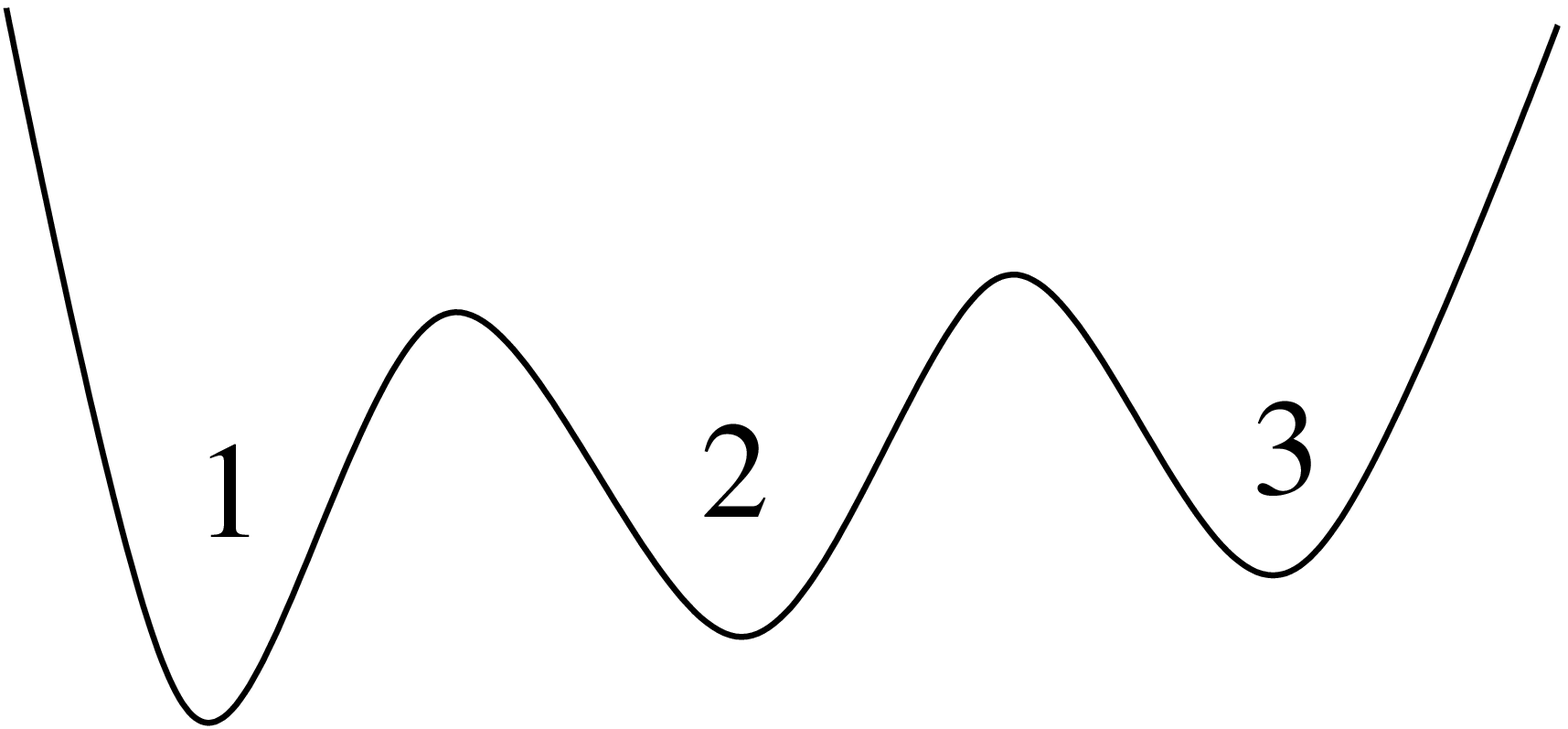} \ \ 
\epsfxsize=5cm\epsffile{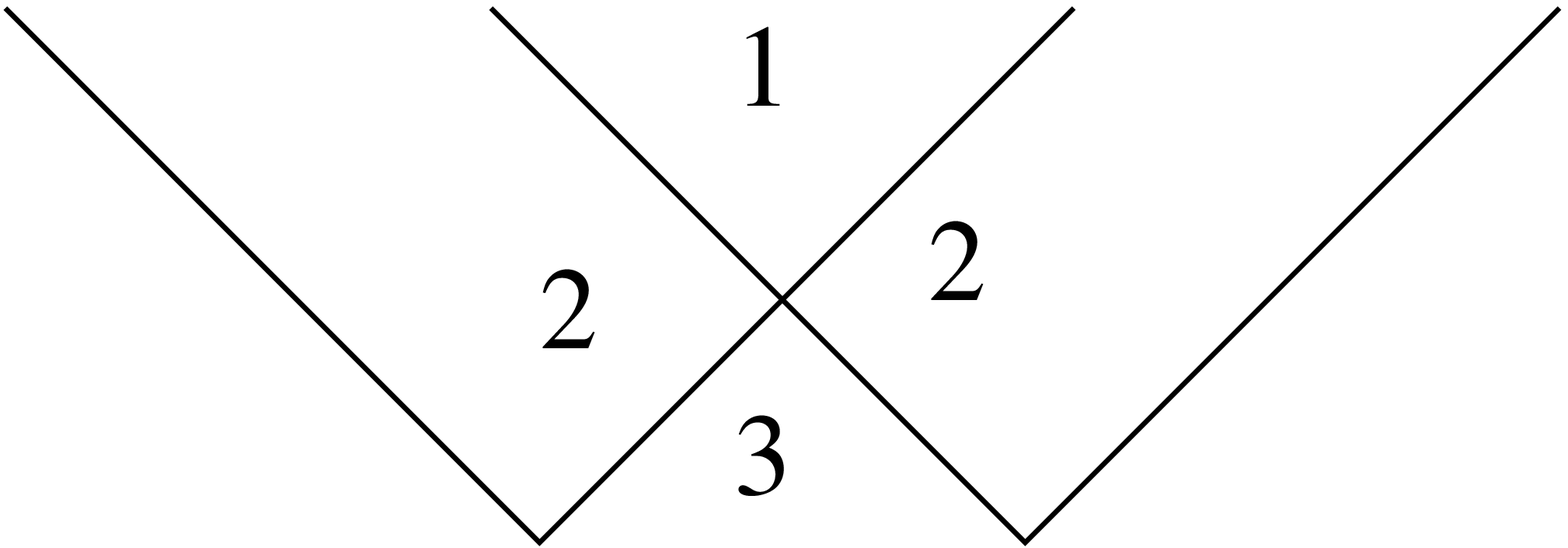} \ \ 
\epsfxsize=5cm\epsffile{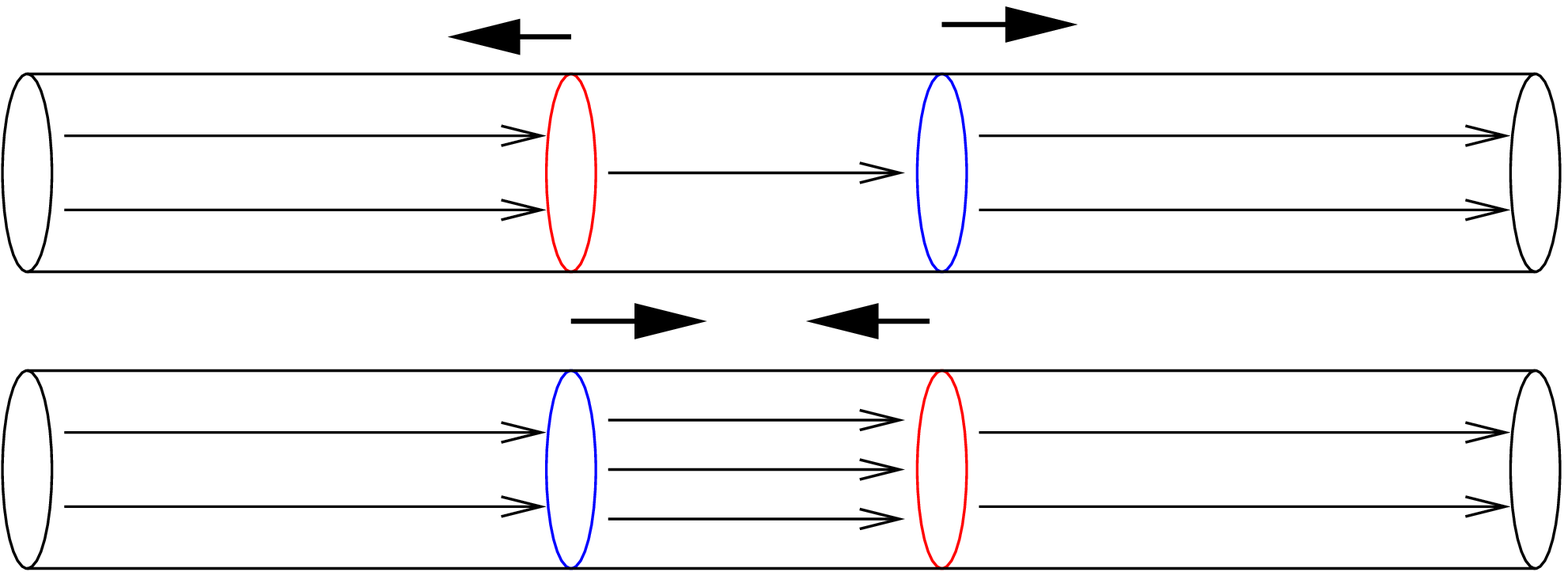}
\caption{A classical transition (middle) can occur when the positions 
of vacua are arranged appropriately (left), as demonstrated in this 
paper.  Alternatively one can have a model like the figure on the right, where 
vacua 1,2,3 have 1,2,3 units of flux and the domain wall is the charged object 
which changes the flux.  If the charged objects freely pass through each 
other, a classical transition is the most natural result of every collision.}
\label{figure:flux}
\end{figure*}

In the context of a scalar field, such a cascade of classical transitions
seems to rely on the existence of roughly evenly spaced vacua in the potential landscape.
Interestingly, in \cite{BlaJos09} it is shown that in a model with
multiple vacua constructed from extra dimensions and fluxes, a classical 
transition is the most natural result, as shown in Figure~\ref{figure:flux}.
In these models, the multiple classical transition structure in 
Figure~\ref{figure:CFT} should be taken seriously.  A similar but simpler 
structure was studied in \cite{FreKle09a} and suggested the possibility of a 
CFT description for an eternal inflating spacetime.

\acknowledgments
We would like to thank Richard Easther, Brian Greene, Matt Johnson, 
Tommy Levi, and especially 
Matt Kleban, Alberto Nicolis and 
Erick Weinberg for many useful discussions.
This work is supported in part by the DOE DE-FG02-92-ER40699,
and by the NASA Astrophysics Theory Program 09-ATP09-0049.
LH thanks members of the CCPP at NYU and the IAS at Princeton
for their fabulous hospitality. EAL thanks the hospitality of Kenyon College where some of this work was done and acknowledge the support of a Foundational Question Institute Mini-Grant.
Research at the Perimeter Institute for Theoretical Physics is supported by the Government of Canada through Industry Canada and by the Province of Ontario through the Ministry of Research \& Innovation.  

\appendix

\section{The Sine-Gordon Example}
\label{sinegordon}

%Notes 09 VI p. 252 - 269.
The sine-Gordon potential is $V = 1 - {\rm cos}\phi$ such that
the equation of motion is $-\partial_t^2 \phi + \partial_x^2 \phi
= {\rm sin}\phi$.
The exact one-soliton solution is
\begin{eqnarray}
\phi = 4 \,{\rm tan}^{-1} {\,\rm exp}
\left[ {x - ut \over \sqrt{1 - u^2}} \right]
\, .
\end{eqnarray}
This interpolates between $\phi = 2\pi$ to the far right
and $\phi = 0$ to the far left.
The exact soliton-anti-soliton pair solution is
\begin{eqnarray}
\phi = 4 \,{\rm tan}^{-1} 
\left[{1\over u} {{\rm sinh} (-ut/\sqrt{1-u^2})
\over {\rm cosh} (x/\sqrt{1-u^2})} \right]
\, .
\end{eqnarray}
This solution describes the collision of an incoming ($0/2\pi$, $2\pi/0$) pair,
resulting in an outgoing ($0/-2\pi$, $-2\pi/0$) pair, {\it irrespective}
of the size of $u$. Note that $u=0$ is \emph{not} a solution -- there exist no stable static solution of more than one sine-Gordon soliton.

\bibliographystyle{apsrev4-1}
\bibliography{all}

\end{document}